\documentclass[aps,prb,floatfix,amsmath,amssymb,twocolumn,showpacs]{revtex4-1}
\usepackage{amsmath, amssymb, graphics, mathrsfs, bm, stackrel,bbold}

\usepackage{subfigure}
\usepackage[usenames,dvipsnames]{xcolor}
\definecolor{Highlight}{rgb}{1,1,0.75}

\allowdisplaybreaks[1]
\usepackage{graphicx}
\usepackage[pdftex, bookmarks={false}, pdfauthor={Rudro Rana Biswas}, pdftitle={Diffusive transport in Weyl semimetals}]{hyperref}
\hypersetup{colorlinks=true, linkcolor=BrickRed, citecolor=Violet, filecolor=OliveGreen, urlcolor=RoyalBlue, filebordercolor={.8 .8 1}, urlbordercolor={.8 .8 0}}
\usepackage[all]{hypcap}
\usepackage{soul}
\setstcolor{Red}

\newcommand\ba{\begin{array}}
\newcommand\ea{\end{array}}
\newcommand\nn{\nonumber}

\newcommand\ri{\right}
\renewcommand\le{\left}

\newcommand{\feyn}[1]{#1\kern-0.45em/}

\renewcommand\a{\alpha}
\newcommand\mba{\mbs{a}}

\renewcommand\c{\psi}

\renewcommand\d{\delta}

\newcommand\D{\Delta}

\newcommand\e{\epsilon}
\newcommand\ve{\varepsilon}

\newcommand\f{\phi}

\newcommand\g{\gamma}

\newcommand\G{\Gamma}

\newcommand\mbJ{\mbs{J}}

\newcommand\mbk{\mbs{k}}

\renewcommand\l{\lambda}
\renewcommand\L{\Lambda}

\newcommand\m{\mu}

\newcommand\p{\pi}
\newcommand\mbp{\mbs{p}}

\newcommand\mbq{\mbs{q}}

\newcommand\rr{\rho}

\newcommand\mbr{\mbs{r}}

\newcommand\s{\sigma}

\newcommand\Ss{\Sigma}
\newcommand\mbss{\mbs{\s}}

\renewcommand\t{\tau}

\renewcommand\th{\theta}

\newcommand\Th{\Theta}
\newcommand\mbv{\mbs{v}}
\newcommand\w{\omega}
\newcommand\W{\Omega}

\newcommand\vx{\chi}

\newcommand\z{\zeta}

\newcommand\imply{\Rightarrow}
\newcommand\grad{\mbs{\nabla}}
\newcommand\la{\langle}
\newcommand\ra{\rangle}
\newcommand\pd{\partial}
\newcommand\mc{\mathcal}
\newcommand\mb{\mathbb}
\newcommand\mbs{\boldsymbol}

\newcommand\ms{\mathscr}

\begin{document}
\title{Diffusive transport in Weyl semimetals}
\author{Rudro R. Biswas}
\email{rrbiswas@illinois.edu}
\affiliation{University of Illinois at Urbana-Champaign, 1110 W.\ Green St., Urbana, IL 61801}
\author{Shinsei Ryu}
\affiliation{University of Illinois at Urbana-Champaign, 1110 W.\ Green St., Urbana, IL 61801}

\begin{abstract}
Diffusion, a ubiquitous phenomenon in nature, is a consequence of particle number conservation and locality, in systems with sufficient damping. In this paper we consider diffusive processes in the bulk of Weyl semimetals, which are exotic quantum materials, recently of considerable interest. In order to do this, we first explicitly implement the analytical scheme by which disorder with anisotropic scattering amplitude is incorporated into the diagrammatic response-function formalism for calculating the `diffuson'. The result thus obtained is consistent with transport coefficients evaluated from the Boltzmann transport equation or the renormalized uniform current vertex calculation, as it should be. We thus demonstrate that the computation of the diffusion coefficient should involve the transport lifetime, and not the quasiparticle lifetime. Using this method, we then calculate the density response function in Weyl semimetals and discover an unconventional diffusion process that is significantly slower than conventional diffusion. This gives rise to relaxation processes that exhibit stretched exponential decay, instead of the usual exponential diffusive relaxation. This result is then explained using a model of thermally excited quasiparticles diffusing with diffusion coefficients which are strongly dependent on their energies. We elucidate the roles of the various energy and time scales involved in this novel process and propose an experiment by which this process may be observed.
\end{abstract}
\pacs{00.00}
\maketitle


%
%

\section{Introduction}

The last decade has witnessed an enormous surge of interest in 2D semimetals, materials with a node in the electronic density of states (DOS) where the valence and conduction bands touch. Stoichiometrically, this band-touching point is also the position of the chemical potential in the undoped state. This undoped state is challenging to implement experimentally because of the two dimensional nature of these materials, which leads to oversize effects due to the environment (e.g, the substrate). Spectacular examples of such 2D semimetals are provided by graphene \cite{2011-geim-fk} and the surface states of strong topological insulators (STI) \cite{2011-hasan-uq}, where the quasiparticles obey the $(2+1)$ dimensional 2-component Dirac equation, thus exhibiting linear dispersion. Since the Dirac equation can have two opposite chiralities and the full lattice band structure should contain quasiparticles with the net chirality of all branches equal to zero, these quasiparticles are required to always come in pairs\cite{1981-nielsen-mz}. In the cases of graphene and STIs these pairs are well-separated, in momentum space and real space respectively. Thus, under appropriate circumstances, their behavior may be understood as arising from the behavior of independent copies of $(2+1)$-D 2-component Dirac fermions. An additional characteristic of these $(2+1)$-D quasiparticles is that there is a relatively easy route for converting them into massive gapped bands and so destroying the critical scale-free nature of the massless theory. This is accomplished by breaking, spontaneously or explicitly, the symmetry that protects the massless character. In graphene this is a combination of time reversal and inversion symmetries\cite{2013-bernevig-qy,2009-drut-uq}; it is the time reversal symmetry in STIs\cite{2010-biswas-uq, 2011-wray-fk}.

In recent years there have been exciting analogous proposals for observing similar phenomena in 3D semimetals\cite{2011-wan-fk, 2011-burkov-fk}. In this case the two component theory near the touching point of the conduction and valence bands is that of Weyl fermions. The theory of these fermions may be obtained by setting the mass term for the relativistic 4-component Dirac equation to zero, and then using one of the pair of decoupled two-component fermions with opposite chiralities that form the 4-component Dirac spinor. After appropriate rotations and rescalings, the Hamiltonian of the theory can always be recast as $\mc{H}_{W} = \pm v \mbss\cdot\mbk$, where $v$ is the velocity, $\mbss\cdot\mbk = \s_{x}k_{x} + \s_{y}k_{y} + \s_{z}k_{z}$ ($\s_{x,y,z}$ are the Pauli spin matrices), $\mbk$ is the deviation from the band touching point in momentum space, and the $\pm$ sign denotes the two possible chiralities allowed. In this case, since the Hamiltonian is a Hermitian $2\times2$ matrix parametrized by three real numbers (apart from an irrelevant shift in the zero energy), the three components of the momentum quench all degrees of freedom and the spectrum cannot be gapped continuously at a single Weyl point. Also, since the Weyl nodes occur in the bulk of a 3D material, they may be less susceptible to being doped by the environment, unlike the 2D Dirac case where such undesired occurrence of (inhomogeneous) doping needs to be actively eliminated by challenging methods like the isolation of the material from its environment (e.g., suspended graphene) or by sensitive chemical control (e.g., in STIs). However, Weyl points must come in pairs of opposite handed-ness and scattering between different Weyl points can destroy the Weyl points, which can happen because of atomic scale disorder. In the absence of such atomic scale disorder, a Weyl point is stable and may be more accessible in the laboratory than a 2D gapless Dirac theory if an appropriate material is found. Another reason why these materials have garnered much interest is that they are expected to have topologically protected gapless surface states which are novel because they have discontinuous Fermi surfaces (Fermi `arcs')\cite{2011-wan-fk}. These are related to the fact that depending on chirality, the node of the theory acts like a source or sink for the Berry curvature flux\cite{2011-balents-fj}. This leads the fermions to exhibit exotic behaviors like the chiral anomaly because of which in the presence of parallel electric and magnetic fields particles are lost or created for Weyl fermions of opposite chiralities\cite{2012-son-yq}.

As in the case of the 2D undoped Dirac theory, the gapless critical nature of the Weyl theory and the chiral nature of the quasiparticles give rise to unconventional features in their transport characteristics and this, in particular, has drawn much research interest\cite{2008-fritz-yq,2012-hosur-uq,2011-burkov-uq}. One aspect has been the effects due to the marginal nature of Coulomb interaction in these theories\cite{2007-sheehy-sv}. Another aspect that attracted much attention in the context of 2D Dirac materials was the possible absence of the Anderson localization transition \cite{2007-nomura-fk} and the nature of associated phenomena, like weak localization effects\cite{2006-mccann-kx}. This discussion originated a couple of decades ago\cite{1986-fradkin-yq,1986-fradkin-rt}, when the effects of quenched disorder in undoped 2D Dirac and 3D Weyl theories were investigated for the cases where the symmetries protecting gaplessness were not broken. The effective coupling induced between the fermions in the replica picture was shown to be marginal in 2D but irrelevant in 3D, in the sense of the Renormalization Group. Thus, the two theories ($2$D and $3$D) are affected very differently by disorder. In the 2D case the presence of a small amount of disorder generates, in a nonperturbative way, a finite DOS at the Dirac node; consequently, the material becomes metallic. In the 3D Weyl case, however, the nodal nature of the Weyl point is preserved unless the disorder strength is larger than a critical value, and so the semimetal phase is stable.

\begin{figure}[h]
\begin{center}
\resizebox{8cm}{!}{\includegraphics{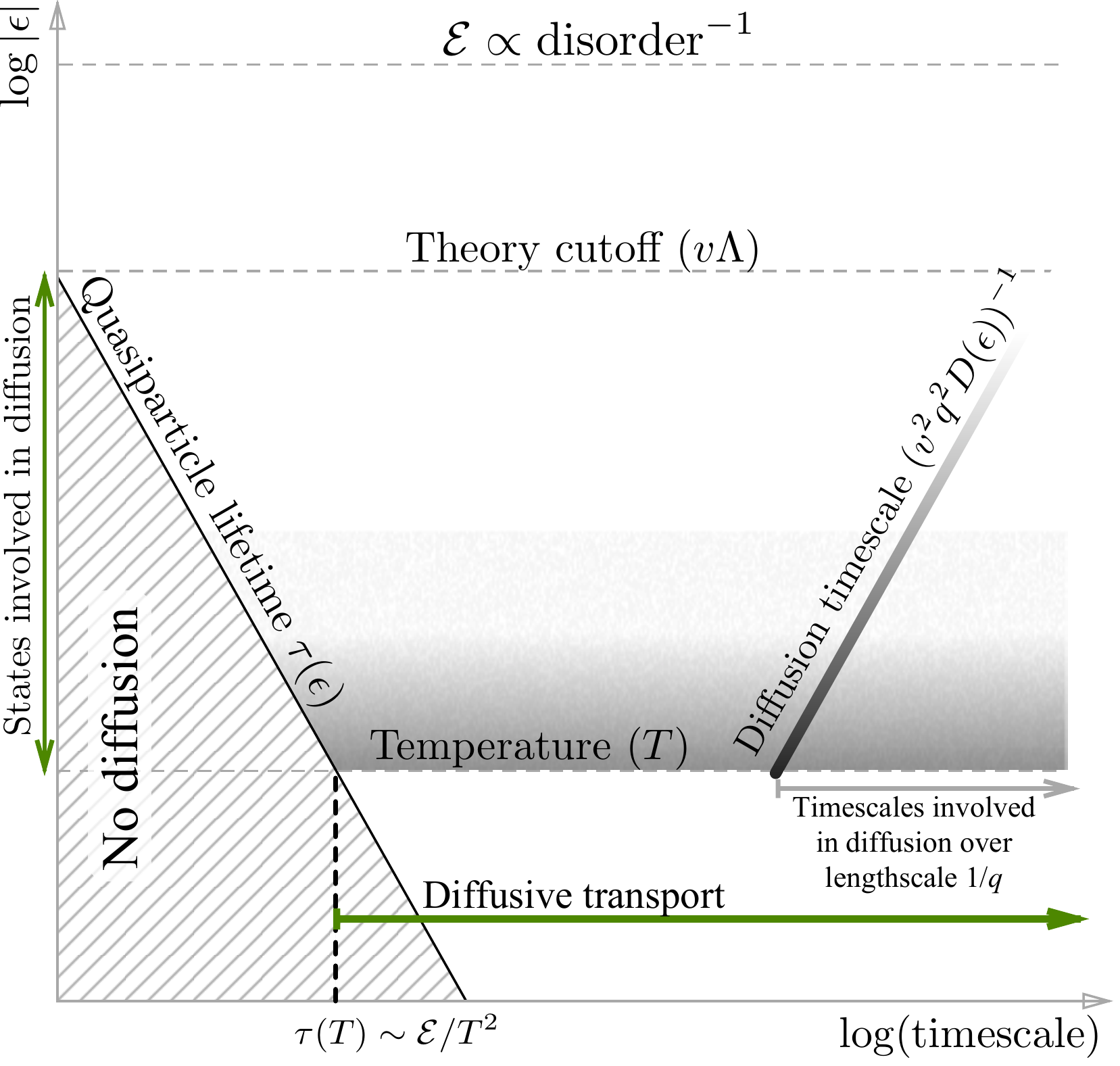}}
\caption{Illustration of the energy and timescales involved in Weyl diffusion. The vertical/horizontal axes are energy/time scales, plotted on a log scale. Quasiparticle (QP) states with energies $|\e|<T$ play a negligible role in diffusion. QPs with energy $\e$ diffuse on timescales much larger than the state lifetime $\t(\e) \propto \e^{-2}$ and their participation is given roughly by the shading (more accurately by the function plotted in Figure~\ref{fig-diffusingenergies}). Thus, diffusion occurs on timescales larger than the lifetime $\t(T)$ of the lowest energy QPs involved. Also, the process, for a given length scale $q^{-1}$, involves a range of diffusing timescales $(v^{2}q^2D(\e))\propto\e^{2}$ attributed to the diffusing QPs at the various energies. Diffusion processes with this wide range of timescales superpose to give an unconventional slow diffusion process shown in Figures~\ref{fig-chargevstime} and \ref{fig-chargeresponsevstime}.}
\label{fig-energyscales}
\end{center}
\end{figure}

In this paper we investigate the existence and idiosyncrasies of diffusive processes in the Weyl semimetal, in the `quantum critical' regime, where the chemical potential is much smaller than the temperature. We do this for weak disorder, assumed to be of the `potential' type and slowly varying, so that there is no scattering between different Weyl nodes which may gap the spectrum. Choosing uncorrelated `vector' disorder (i.e terms proportional to the Pauli matrices in the Hamiltonian) is found to not change the underlying physics. Its effect on diffusion in the isotropic random case will be shown to be a modification of the ratio of the transport to the quasiparticle lifetimes. Unlike the case of the 2D Dirac fermions, for which the effect of disorder on the nodal states is nonperturbative, in the case of the 3D Weyl semimetal disorder is irrelevant and thus well-known perturbative techniques may be used\cite{1961-baym-yq,1985-lee-uq,2010-altland-vn}. However, there are a handful of nuanced differences from the use of these methods for a conventional Fermi gas.

First, the scattering from potential disorder is highly anisotropic here and is characterized by a complete suppression of exact backscattering\cite{1998-ando-lr, 1998-ando-ly}. This is dramatically different from the textbook example of a simple Fermi gas and is known from the Boltzmann approach, or the calculation of the renormalized uniform current vertex, to result in the transport lifetime replacing the state lifetime in the Drude expression of the conductivity\cite{2004-bruus-lq,1958-edwards-fk}. However, this result has not been explicitly derived before for the density response function in the diagrammatic approach, which is used to calculate quantum effects beyond the Boltzmann approach (e.g., weak localization). We shall reconcile these two approaches, demonstrating how to calculate the charge vertex renormalization in the `conserving approximation'\cite{1961-baym-yq} for disorder (the `diffuson'), with an arbitrary anisotropic scattering amplitude. We thus show that the result obtained is consistent with the Boltzmann picture. We find that for disorder with a potential character, the transport time is $3/2$ times the state lifetime (Equation~\eqref{eq-tautrscalardisorder}). The inclusion of isotropic vector disorder reduces this ratio to a number between $3/2$ and $9/10$ (Equation~\eqref{eq-tautrvectordisorder}).

The second characteristic of the diffusion process in Weyl semimetals is that there is a node in the density of states (DOS) at zero energy and the DOS increases linearly with the energy. Due to the node in the DOS, the quasiparticle lifetime becomes infinite at the Fermi level (zero energy) and one expects that at zero temperature transport will be ballistic (however, we can naively expect it to be zero since the DOS is zero at the node). The situation is quite different at a finite temperature $T$ for a sufficiently large sample. We argue that when the disorder is weak, the transport process may be modeled as diffusion by particle-hole pairs with an energy-dependent diffusion coefficient, which becomes smaller at higher energies. The contribution of these pairs to transport is, however, not uniform but proportional to the product of the DOS and the slope of the Fermi distribution function, which is greatly suppressed at energies less that $T$ (see Figure~\ref{fig-diffusingenergies}). As a result, the contribution of the ballistic zero energy modes may be neglected self-consistently. The bulk of diffusive transport occurs via modes with excitation energies that are of the order of or greater than $T$. Since the quasiparticle lifetime is a decreasing function of energy, this means that the diffusive process kicks in for timescales larger than the state lifetime for quasiparticles at energy $T$. Also, since higher energy quasiparticles have a smaller diffusion coefficient and diffuse slowly, this combined diffusive process is shown to result in diffusive relaxation of particle density, which is qualitatively different from and slower than what we obtain for a diffusive process with a single diffusion coefficient. While the latter (conventional) process exhibits relaxation that exponentially decays in the time $t$, in a Weyl semimetal the relaxation process is exponential in $t^{1/3}$ -- a stretched exponential in $t$. Hence it is much slower (Equations \eqref{eq-chargeresponsetimedomain} and \eqref{eq-chargeresponse4}; Figures~\ref{fig-chargevstime} and \ref{fig-chargeresponsevstime}). A process such as this is described by diffusion with a memory function\cite{1990-forster-qq} which decays slowly. There does not exist a well-defined timescale beyond which it may be neglected. The energy and time scales involved in Weyl diffusion are summarized in Figure~\ref{fig-energyscales}.

A consequence of the finite temperature diffusion is that the Weyl semimetal has a finite DC conductance, which at zero temperature may naively be expected to be zero, since the DOS is zero at the Fermi point. This finite temperature DC conductance, obtained when only impurity scattering is present, has been derived previously, except for the crucial substitution of the transport lifetime by the quasiparticle lifetime. This requires the gauge invariance abiding current vertex renormalization, which was ignored in those works.

The remainder of this paper is organized as follows. We first introduce the phenomenon of diffusion in the presence of a general memory function, in section~\ref{sec-diffusionintro}. We then explain our technique for treating the case of quenched disorder with anisotropic scattering when calculating the density vertex renormalization in the diffuson approximation, in section~\ref{sec-anisotropicmethod}. Proceeding to the case of Weyl semimetals in section~\ref{sec-weyldiffusion}, we first calculate the electron self energy in the self consistent Born approximation (SCBA) and show that it predicts a semimetal to metal phase transition, as disorder strength is increased\cite{1986-fradkin-yq,1986-fradkin-rt}. We subsequently only consider the case of weak disorder when the Born approximation is sufficient. The density response function is then calculated using the techniques from the previous section, following which we elaborate on the idiosyncrasies arising from the nodal nature and the strong variation of the DOS with energy. We conclude this section with a phenomenological model of diffusion where independently diffusing thermally excited quasiparticles with an energy-dependent diffusion coefficient and show that it is equivalent to the results obtained for the Weyl semimetal, thus clarifying the physical picture that is presented in this paper. In section~\ref{sec-leftover} we comment on the DC conductivity in Weyl semimetals and the roles of vector disorder and Coulomb interactions. In conclusion, we propose an experiment (see Figure~\ref{fig-exptsetup}) using currently available techniques to observe this novel slow diffusive relaxation when a Weyl semimetal is found or fabricated in the laboratory.

\section{Diffusion and the density response function}
\label{sec-diffusionintro}

Diffusion is a consequence of particle number conservation. The rate of change of the number density is, to the first approximation, related linearly to its history via a diffusion memory function\cite{1990-forster-qq} $M$ (assuming space and time translation invariance on the average; we shall always work in the reciprocal $\mbq$-space):
\begin{align}\label{eq-diffusionmemorydef0}
\pd_{t}n_{\mbq}(t) + \int_{0}^{\infty}dt'\,M_{\mbq}(t')n_{\mbq}(t-t') &= 0
\end{align}
The local nature of the underlying microscopic theory, in combination with the conserved nature of particle number, ensures that $\hat{M}_{\mbq}(\W) \to 0$ as $\mbq\to 0$. For long enough timescales when all low frequency `reactive' modes are damped out and the system enters the `diffusive' regime, $\hat{M}_{\mbq}(\W) \propto q^{2}$ as $\mbq\to 0$.
The generalized Laplace transform\footnote{The generalized Laplace transform of a function $f(t)$ is defined via $$\tilde{f}(z) = \int_{-\infty}^{\infty} \frac{d\w}{2\p}\frac{\hat{f}(\w)}{z - \w},$$ where $\hat{f}(\w)$ is the usual Fourier transform of $f(t)$. With this definition, $\tilde{f}(\w\pm i 0)$ is the Fourier transform of $\mp i\Th(\pm t)f(t)$, $\Th$ denoting the Heaviside step function.} (denoted by a tilde over the function name) providing the $t>0$ evolution of this generalized diffusion equation (thus Im$(z)>0$ below and throughout this paper) in terms of the initial density $n_{\mbq}(0)$ at $t=0$ is given by:
\begin{align}\label{eq-diffusionmemorydef1}
\tilde{n}_{\mbq}(z) &= \frac{n_{\mbq}(0)}{z - \tilde{M}_{\mbq}(z)}, \qquad \text{Im}(z)>0
\end{align}
If there is a timescale $\t_{M}$, decided by microscopic processes, beyond which $M$ is negligible as the system loses its memory, then on timescales much longer than $\t_{M}$ Equations~\eqref{eq-diffusionmemorydef0} and  above reduces to the well-known `Markovian' diffusion equation
\begin{align}\label{eq-diffusionequationmaster}
(\pd_{t} + D q^{2}) n_{\mbq}(t) &= 0
\end{align}
and so
\begin{subequations}\label{eq-diffusiondecay}
\begin{align}
\tilde{n}_{\mbq}(z) &= \frac{n_{\mbq}(0)}{z + i D q^{2}},\quad |z|\t_{M} \ll 1\\
\imply n_{\mbq}(t) &= n_{\mbq}(0)e^{-Dq^{2}t}, \quad t\gg\t_{M}
\end{align}
\end{subequations}
The diffusion coefficient $D$ is defined through
\begin{align}\label{eq-diffconstmemoryfunction}
D &= \int_{0}^{\approx \t_{M}}dt'\,\frac{M_{\mbq}(t')}{q^{2}} \simeq \le. i \frac{\tilde{M}_{\mbq}(z)}{q^{2}}\ri|_{\text{Im}(z)>0,|z|\t_{M} \ll 1}
\end{align}
However, in the absence of such a well-defined timescale $\t_{M}$, as will be the case with Weyl quasiparticles below, we need to use the general form Equation~\eqref{eq-diffusionmemorydef0} or \eqref{eq-diffusionmemorydef1}. Given this linear relation between $\tilde{n}_{\mbq}(z)$ and $n_{\mbq}(0)$, we can deduce the low frequency behavior of the density linear response function $\tilde{\vx}_{\mbq}(z)$. This response function $\vx_{\mbq}$ quantifies the density response to a chemical potential wave $\m_{\mbq}(t)$ according to:
\begin{align}\label{eq-densityresponsefunction1}
n_{\mbq}(t) &= -i\int_{-\infty}^{t}dt'\, \vx_{\mbq}(t-t')\m_{\mbq}(t')\nn\\
\imply\hat{n}_{\mbq}(\W)&= \tilde{\vx}_{\mbq}(\W+i0)\hat{\m}_{\mbq}(\W)
\end{align}
where $\hat{f}$ denotes the usual temporal Fourier transform of $f(t)$. At low frequencies $z$ with Im$(z)>0$, we can show that the frequency dependence of $\tilde{\vx}_{\mbq}(z)$ is completely determined by the memory function $\tilde{M}_{\mbq}(z)$ via the relation\cite{1990-forster-qq}
\begin{align}\label{eq-chidiffusionpole}
\tilde{\vx}_{\mbq}(z) &= \le(\frac{- \tilde{M}_{\mbq}(z)}{z -\tilde{M}_{\mbq}(z)}\ri)\tilde{\vx}_{\mbq}(i0+)\\
&\approx \le(\frac{ i D q^{2}}{z + i D q^{2}}\ri)\tilde{\vx}_{\mbq}(i0+), \quad |z|\t_{M}\ll 1\nn
\end{align}

\section{Electron diffusion when the impurity scattering amplitude is anisotropic}
\label{sec-anisotropicmethod}

Before embarking on finding the diffusion mechanism in Weyl semimetals, we shall take a detour and calculate the charge density response in an electron gas in the presence of random potential disorder, when the scattering amplitude is \emph{anisotropic}, i.e, varies with the angle between incoming and outgoing momenta. It is well known from the Boltzmann transport equation approach or a diagrammatic calculation of the renormalized uniform current vertex that this changes the appropriate microscopic timescale entering the diffusion process from the quantum state lifetime $\t$ to the transport lifetime $\t^{\text{tr}}$, both of which are defined in Equations~\eqref{eq-selfenergy} and \eqref{eq-transportlifetime} below. Since the diagrammatic method is used for deriving quantum corrections to the Boltzmann result (like weak localization) it is useful to be able to treat the case of anisotropic scattering in this framework also. Even though this seems to be fundamental enough to have been derived before, we found the research literature to be lacking in this respect and shall derive the procedure in this section.

\subsection{Electron self energy $\Ss$}

We first recount the well-known procedure to obtain the single particle Green's function in the presence of a static disorder potential $U$, which has zero mean ($\le\la U(\mbr)\ri\ra = 0$) and satisfies the `white noise' criterion at long enough length scales:
\begin{align}\label{eq-whitenoise}
\le\la U(\mbr)U(\mbr')\ri\ra &= \z\, \d(\mbr-\mbr')
\end{align}
Here $\le\la \bullet \ri\ra$ denotes averaging over different realizations of disorder. As an example, if we consider a density $n_{\text{imp}}$ of random short range potential wells $V$ with $\le\la V(\mbr)\ri\ra = 0$ and $\le\la \le(\int V(\mbr) d^{d}r\ri)^{2}\ri\ra = U_{0}^{2}$, where $d$ is the spatial dimension, then we can show that $\z = n_{\text{imp}}  U_{0}^{2}$. We shall assume that the electron wave functions are labelled by their momentum $\mbk$ and that the dispersion is isotropic (for convenience of calculation) and is given by $\ve_{\mbk} \equiv \ve_{k}$, and that the scattering amplitude $\le\la \mbk \le|U \ri| \mbq\ri\ra \equiv U_{\mbk\mbq}$ is allowed to depend sensitively on the angle $\th_{\mbk\mbq}$ between the incoming and outgoing momenta. This is true, for example, in the cases of Dirac or Weyl fermions, the latter of which we shall consider in the next section. In that case we can assume the following general form for the disorder-averaged squared magnitude of the scattering amplitude involving states at similar energies\footnote{Scattering processes will effectively mix states whose energies are separated by the inverse state lifetime, i.e, the energy `linewidth'.}:
\begin{align}
\frac{1}{\ms{V}}\le\la \le|U_{\mbk\mbq}\ri|^{2}\ri\ra = \mc{U}(\e,\th_{\mbk\mbq}), \; \e = \ve_{\mbk} \approx \ve_{\mbq}
\end{align}
where $\ms{V}$ is the total volume\footnote{This arises from the delta function in momentum space $ \bigl< U_{\mbk\mbq}U_{\mbp\mbs{l}}\bigr> \propto \d(\mbk+\mbp - \mbq - \mbs{l}) $.}. We shall assume for our purposes that the dependence on $\e$ is mild (defined by a relation analogous to \eqref{eq-slowdos}) and the above form will be substituted in place of the impurity-correlation potential denoted by the starred vertex dashed line in Feynman diagrams.

\begin{figure}[h]
\begin{center}
\resizebox{6cm}{!}{\includegraphics{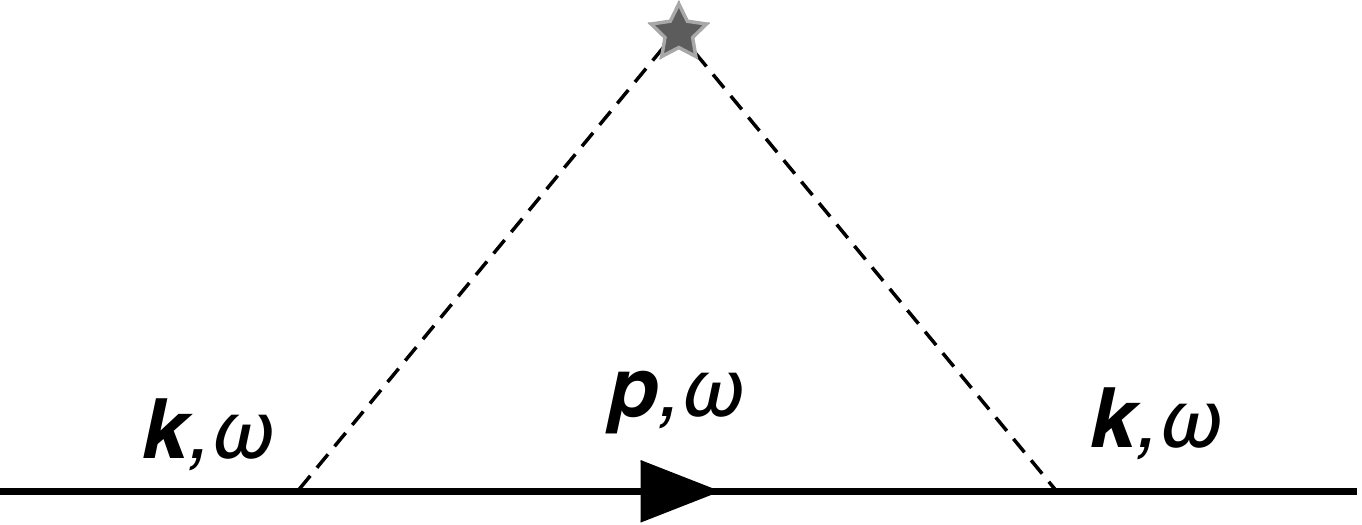}}
\caption{The self energy $\Ss(\w,\mbk)$ in the Born approximation.}
\label{fig-selfenergyborn}
\end{center}
\end{figure}

The disorder-averaged retarded/advanced electronic self energy $\Ss^{R/A}(\w,\mbk)$ is calculated, in the Born approximation, by the Feynman diagram shown in Figure~\ref{fig-selfenergyborn}. This yields the effective state lifetime $\t(\w)$ to be, for $\w \approx \ve_{\mbk}$,
\begin{align}\label{eq-selfenergy}
\text{Im}(\Ss^{R/A}(\w,\mbk)) &= \text{Im}\le[\int \frac{d^{d}p}{(2\p)^{d}}\frac{\ms{V}^{-1}\le\la |U_{\mbk\mbp}|^{2}\ri\ra}{\w - \ve_{\mbp} \pm i0}\ri]\nn\\
&\!\!\!\!= \mp \p g(\w)\le\la \mc{U}(\w, \th_{\mbk\mbp})\ri\ra_{\hat{\mbp}} \equiv \mp \frac{1}{2\t(\w)}
\end{align}
Here, $g(\e)$ is the electronic density of states (DOS) at energy $\e$ and $\le\la \bullet\ri\ra_{\hat{\mbp}}$ averages the enclosed expression over directions of $\mbp$. For example,
\begin{align}
\le\la \mc{U}(\w, \th_{\mbk\mbp})\ri\ra_{\hat{\mbp}} &\stackrel{d = 2}{=} \int_{0}^{2\p} \frac{d\th_{\mbk\mbp}}{2\p}\,\mc{U}(\w, \th_{\mbk\mbp})\nn\\
&\stackrel{d = 3}{=} \int_{0}^{\p} \frac{\sin\th_{\mbk\mbp}\,d\th_{\mbk\mbp}}{2}\,\mc{U}(\w, \th_{\mbk\mbp})
\end{align}
The real part of $\Ss$ renormalize the spectrum and quasiparticle weight, both of which are not going to affect the following discussion and will be neglected in this paper. The `quasiparticle' peak of the single particle Green's function is thus going to be given by
\begin{align}\label{eq-greensfunctiondiagonalform}
G^{R/A}(\w,\mbk) &\simeq \frac{1}{\w - \ve_{\mbk} \pm \frac{i}{2\t(\w)}}
\end{align}

\subsection{The density vertex correction}

\begin{figure}[h]
\begin{center}
\resizebox{8cm}{!}{\includegraphics{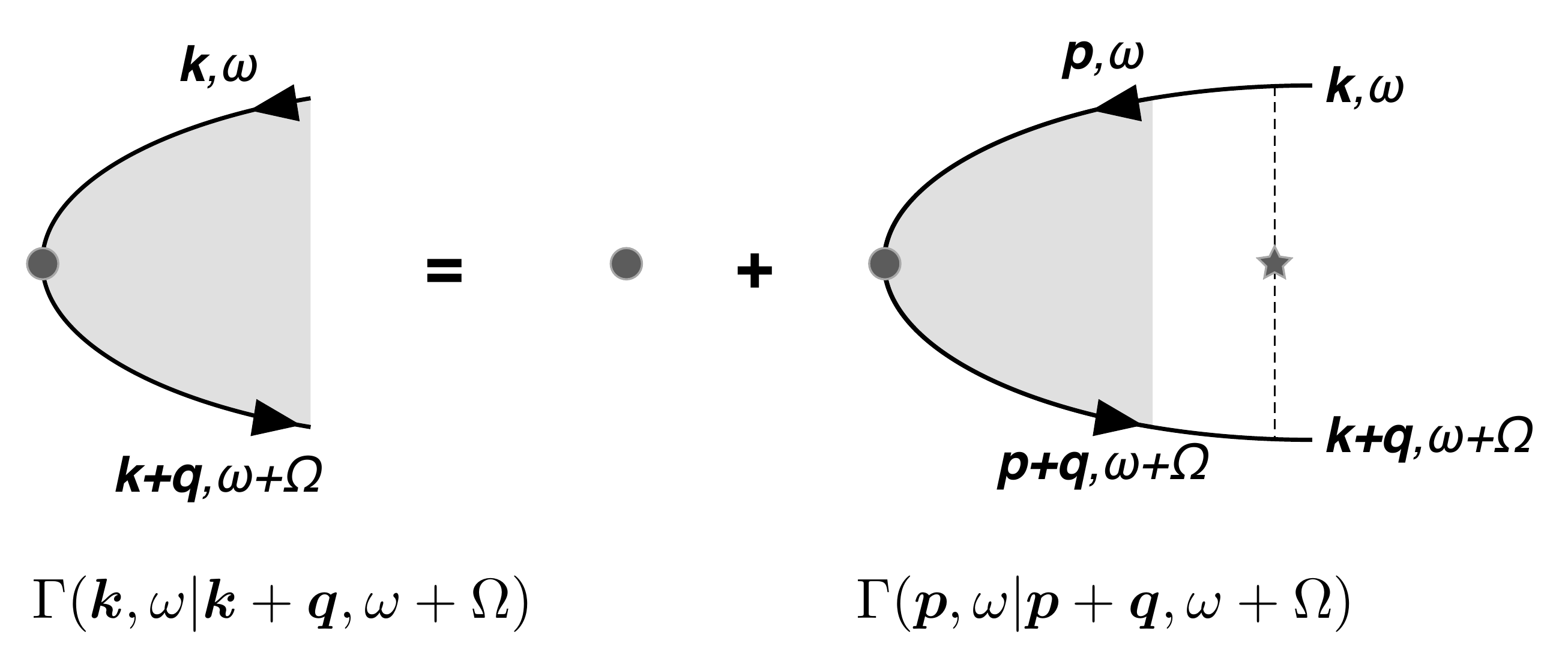}}
\caption{Bethe-Salpeter equation for the vertex $\hat{\rr}_{\mbq}(\W)$.}
\label{fig-vertexequation}
\end{center}
\end{figure}
The diffusion pole that is expected to arise in the density response function Eq.~\eqref{eq-chidiffusionpole} is obtained by correctly calculating the density vertex, maintaining number conservation\cite{1961-baym-yq}. For this, we need to solve the Bethe-Salpeter equations for the density vertex\cite{2010-altland-vn, 1958-edwards-fk, 1980-vollhardt-fk, 1963-abrikosov-ly} corresponding to $\hat{\rr}_{\mbq}(\W)$ (as is schematically sketched in Figure~\ref{fig-vertexequation}):

\begin{widetext}
\begin{align}\label{eq-vertexequation}
&\G(\mbk, \w |\mbk+\mbq, \w+\W) = 1 + \int \frac{d^{d}p}{(2\p)^{d}} \frac{\le\la \le|U_{\mbk\mbp}\ri|^{2}\ri\ra}{\ms{V}}\G(\mbp, \w |\mbp+\mbq, \w+\W) G^{A}(\w, \mbp)G^{R}(\w+\W, \mbp+\mbq)
\end{align}
\end{widetext}
There are also contributions corresponding to the products $G^{R}G^{R}$ and $G^{A}G^{A}$ which we shall neglect in this paper as they do not influence the low frequency diffusive contribution to transport\cite{2004-bruus-lq} and only serve to cancel out unphysical contributions from the high energy states. We shall also only be interested in values of $\w \approx \ve_{\mbk}$, which tells us that the values of $\mbp$ that contribute substantially to the integral on the right are those for which $\ve_{\mbp}\approx\w\approx\ve_{\mbk}$ (`classical' ballistic transport in between collisions). Thus the $\mbk$-dependence of $\G$ is reduced to those through $\ve_{\mbk}$ and the dependence on the orientation of $\mbk$ w.r.t $\mbq$, i.e via the angle $\th_{\mbk\mbq}$. In the following, we shall thus write the vertex renormalization as $\G(\ve_{\mbk}, \th_{\mbk\mbq}, q, \W) \equiv \G(\mbk, \w |\mbk+\mbq, \w+\W)$ when $\w\approx \ve_{\mbk}$. Equation~\eqref{eq-vertexequation} then becomes
\begin{widetext}
\begin{align}\label{eq-vertexequation2}
&\G(\ve_{\mbk}, \th_{\mbk\mbq}, q, \W) = 1 + \int \frac{d^{d}p}{(2\p)^{d}} \mc{U}(\ve_{\mbk}, \th_{\mbk\mbp})\G(\ve_{\mbp}, \th_{\mbp\mbq}, q, \W) G^{A}(\ve_{\mbk}, \mbp)G^{R}(\ve_{\mbk}+\W, \mbp+\mbq)
\end{align}
\end{widetext}
As far as we can tell, this explicit dependence on $\th_{\mbk\mbq}$ has been neglected in previous published research. This is justified only when the impurity scattering is independent of the angle between the incoming and outgoing momenta. Because of this, past calculations involving the vertex renormalization have failed to distinguish between the transport lifetime and state lifetime, and cannot be applied to the cases of Dirac and Weyl fermions without invoking the Boltzmann transport equation.

In order to solve Equation~\eqref{eq-vertexequation2}, we decompose the angular functions into a complete basis of angular modes for $\th_{\mbk\mbq}$, the specifics of which depend on the dimension. For $d=2$, these would be the cosines $\cos(n\th)$ while in $d=3$, which we shall concentrate on, we shall use the Legendre polynomials $P_{\ell}(\cos\th)$ which are the $m=0$ components of the spherical harmonics $Y_{\ell m}(\th,\f)$ and are independent of the azimuthal angle $\f$. Thus we define
\begin{align}\label{eq-vertexdecomposition}
\G(\ve_{\mbk}, \th_{\mbk\mbq}, q, \W) &= \sum_{n=0}^{\infty} \g_{n}(\ve_{\mbk},q, \W)P_{n}(\cos\th_{\mbk\mbq})
\end{align}
and also the angular averages
\begin{align}\label{eq-Uangularavg}
\le\la \mc{U}(\e, \th) P_{n}(\cos\th)\ri\ra_{\text{ang}} &= \frac{\a_{n}}{2\p g(\e)\t(\e)}
\end{align}
In particular, using Equation~\eqref{eq-selfenergy} we see that
\begin{align}
\a_{0} = 1
\end{align}
Also, we can define the transport lifetime via
\begin{align}\label{eq-transportlifetime}
\frac{1}{\t^{\text{tr}}(\ve_{\mbk})} &= 2\p \int \frac{d^{d}p}{(2\p)^{d}} \mc{U}(\ve_{\mbk}, \th_{\mbk\mbp})(1 - \cos\th_{\mbk\mbp})\d(\ve_{\mbp} - \ve_{\mbk})\nn\\
&= \frac{1 - \a_{1}}{\t(\ve_{\mbk})}
\end{align}

Using the shorthand notations $\th_{\mba} \equiv \th_{\mba\mbk}$ for any vector $\mba$, $\ve_{\mbk} = \e$, $\t \equiv \t(\ve_{\mbk})$, $\D = \e - \ve_{\mbp} + \frac{i}{2\t}$ and $\d = \mbv_{\mbp}\cdot\mbq - \W = v q \cos\th_{\mbp\mbq} - \W$, we can manipulate the Bethe-Salpeter equation~\eqref{eq-vertexequation2} as follows:

\begin{align}\label{eq-vertexequation3}
&\G(\th_{\mbq},q,\W) - 1\nn\\
&= \int \frac{d^{d}p}{(2\p)^{d}} \mc{U}(\e, \th_{\mbp})\G(\e,\th_{\mbq\mbp},q,\W)G^{A}(\e,\mbp)G^{R}(\e+\W,\mbp+\mbq)\nn\\
&=  \int \frac{d^{d}p}{(2\p)^{d}} \G(\e,\th_{\mbq\mbp},q,\W)\frac{\mc{U}(\e, \th_{\mbp})}{\D^{*}(\D-\d)}\nn\\
&= \int d\ve_{\mbp} g(\ve_{\mbp}) \bigg\la\G(\e,\th_{\mbq\mbp},q,\W)\frac{\mc{U}(\e, \th_{\mbp})}{|\D|^{2}}\nn\\
&\qquad\qquad\qquad\quad\times \le(1 + \frac{\D^{*} \d}{|\D|^{2}} + \frac{\d^{2}(\D^{*})^{2}}{|\D|^{4}} + \ldots\ri)\bigg\ra_{\hat{\mbp}}\nn\\
&\equiv \ms{T}_{1} + \ms{T}_{2} + \ms{T}_{3} + \ldots
\end{align}
This expansion is justified when $|\d|\ll |\D|$, i.e.
\begin{align}\label{eq-smallparameters1}
|\W|, vq \ll \frac{1}{\t(\ve_{\mbp})}\approx \frac{1}{\t}
\end{align}
We have expanded till the third term to obtain the lowest order contributions from both $\W$ and $q$, as will be shown below. These terms are evaluated below.

\subsubsection{Evaluating $\ms{T}_{1}$}

Using the decomposition in Equation~\eqref{eq-vertexdecomposition} and suppressing the arguments of $\g_{n}$, we obtain
\begin{align}\label{eq-T1initial}
\ms{T}_{1}&= \sum_{n} \g_{n}\int d\ve_{\mbp} \frac{g(\ve_{\mbp})}{|\D|^{2}}\le\la \mc{U}(\e, \th_{\mbp\mbk})P_{n}(\cos\th_{\mbq\mbp})\ri\ra_{\hat{\mbp}}
\end{align}
Here, we shall make the assumption that the DOS $g(\e)$ and $\mc{U}$ are slow functions of $\e$ over the scale of $\t^{-1}$
\begin{align}\label{eq-slowdos}
\frac{1}{\t(\e)}\frac{g'(\e)}{g(\e)} \ll 1
\end{align}
which allows us to also replace $g(\ve_{\mbp}) \to g(\e)$ in  the energy integral in Equation~\eqref{eq-T1initial} and perform the remaining Lorentzian integral
\begin{align}\label{eq-gg1eval}
\int  \frac{d\ve_{\mbp}}{|\D|^{2}} &= 2\p\t(\e)
\end{align}
to obtain
\begin{align}
\ms{T}_{1}&=2\p g(\e)\t(\e) \sum_{n} \g_{n}\le\la \mc{U}(\e, \th_{\mbp\mbk})P_{n}(\cos\th_{\mbq\mbp})\ri\ra_{\hat{\mbp}}
\end{align}
Now, we can show\footnote{Using the formula \url{http://dlmf.nist.gov/18.18#E9}.} that angular averaging $\le\la \bullet \ri\ra_{\mbp\perp\mbk}$ over the azimuthal angle of $\mbp$, when $\mbk$ provides the zenith direction, yields
\begin{align}\label{eq-legendreavg}
\le\la P_{n}(\cos\th_{\mbq\mbp}) \ri\ra_{\mbp\perp\mbk} &= P_{n}(\cos\th_{\mbq})P_{n}(\cos\th_{\mbp})
\end{align}
which helps us simplify further and use Equation~\eqref{eq-Uangularavg} to obtain the result
\begin{align}\label{eq-T1final}
\ms{T}_{1}&= \sum_{n} \g_{n} \a_{n} P_{n}(\cos\th_{\mbq})
\end{align}

\subsubsection{Evaluating $\ms{T}_{2}$}

Using the decomposition in Equation~\eqref{eq-vertexdecomposition} we obtain
\begin{align}\label{eq-T2initial}
\ms{T}_{2}&= \sum_{n} \g_{n}\int d\ve_{\mbp} \frac{g(\ve_{\mbp})\D^{*}}{|\D|^{4}}\times\nn\\
&\qquad\quad\le\la \mc{U}(\e, \th_{\mbp\mbk})P_{n}(\cos\th_{\mbq\mbp})\le(v q \cos \th_{\mbq\mbp} - \W\ri)\ri\ra_{\hat{\mbp}}
\end{align}
As before, the DOS may be replaced by $g(\e)$ and the energy integral evaluates to
\begin{align}
\int d\ve_{\mbp} \frac{\D^{*}}{|\D|^{4}} &= - 2 i \p \t^{2}
\end{align}
In order to evaluate the angular integral, we need to use the relation
\begin{align}\label{eq-legendreprop1}
x P_{n}(x) &= \frac{(n+1)P_{n+1}(x) + n P_{n-1}(x)}{2n+1}\nn\\
&\equiv d_{n}P_{n+1} + f_{n}P_{n-1}
\end{align}
and the angular averaging property Equation~\eqref{eq-legendreavg} to obtain
\begin{align}\label{eq-T2final}
\ms{T}_{2}&= - i \t \sum_{n} \g_{n} \big[\W P_{n}(\th_{\mbq})\a_{n} \nn\\
&- vq(d_{n}P_{n+1}(\th_{\mbq})\a_{n+1} + f_{n}P_{n-1}(\th_{\mbq}) \a_{n-1})\big]
\end{align}

\subsubsection{Evaluating $\ms{T}_{3}$}

Using the decomposition in Equation~\eqref{eq-vertexdecomposition} we obtain
\begin{align}\label{eq-T3initial}
\ms{T}_{3}&= \sum_{n} \g_{n}\int d\ve_{\mbp} \frac{g(\ve_{\mbp})(\D^{*})^{2}}{|\D|^{6}}\times\nn\\
&\qquad\le\la \mc{U}(\e, \th_{\mbp\mbk})P_{n}(\cos\th_{\mbq\mbp})\le(v q \cos \th_{\mbq\mbp} - \W\ri)^{2}\ri\ra_{\hat{\mbp}}
\end{align}
Extracting the DOS from the energy integral, the remaining energy integral reduces to
\begin{align}
\int d\ve_{\mbp} \frac{(\D^{*})^{2}}{|\D|^{6}} &= - 2 \p \t^{3}
\end{align}
In order to evaluate the angular integral in this case, we need to use the relation
\begin{align}
x^{2}P_{n} &= a_{n}P_{n+2} + b_{n}P_{n} +c_{n}P_{n-2}
\end{align}
where
\begin{subequations}
\begin{align}
a_{n} &= \frac{(n+1)(n+2)}{(2n+1)(2n+3)}\\
b_{n} &= \frac{4n^{3} + 6n^{2}-1}{(2n+3)(2n+1)(2n-1)}\\
c_{n} &= \frac{n(n-1)}{(2n+1)(2n-1)}
\end{align}
\end{subequations}
Using these, Equation~\eqref{eq-legendreprop1} and the angular averaging property Equation~\eqref{eq-legendreavg} in Equation~\eqref{eq-T3initial}, we find
\begin{align}\label{eq-T3final}
\ms{T}_{3} &= - \t^{2}\sum_{n}\g_{n}\bigg[\W^{2} P_{n}(\th_{\mbq})\a_{n} +\nn\\
&\!\!\!\!v^{2} q^{2} \le(a_{n}P_{n+2}(\th_{\mbq})\a_{n+2} + b_{n}P_{n}(\th_{\mbq})\a_{n}+c_{n}P_{n-2}(\th_{\mbq})\a_{n-2}\ri) \nn\\
& - 2 \W v q (d_{n}P_{n+1}(\th_{\mbq})\a_{n+1} + f_{n}P_{n-1}(\th_{\mbq}) \a_{n-1})\bigg]
\end{align}

\subsubsection{Recursion relations for $\g_{n}$}

Substituting Equations~\eqref{eq-T1final}, \eqref{eq-T2final} and \eqref{eq-T3final} in Equation~\eqref{eq-vertexequation3}, using the decomposition \eqref{eq-vertexdecomposition} and comparing the coefficients of $P_{n}(\th_{\mbq})$ on either side, we obtain a set of recursion relations:
\begin{align}\label{eq-recursion}
\g_{n} - \d_{n0} &= \a_{n}\Big[ \g_{n}\le(1 + i\t\W - \t^{2}\le(b_{n}v^{2}q^{2} + \W^{2}\ri)\ri)\nn\\
& - \t^{2} v^{2}q^{2}(\g_{n-2}a_{n-2} + \g_{n+2}c_{n+2})\nn\\
&\!\!\!\!\!\!\!\!\!\!\!\!\!\!\!\! + \le(-i\t v q  + 2\W vq\t^{2}\ri)\le(\g_{n+1}f_{n+1}+ \g_{n-1}d_{n-1}\ri)\Big]
\end{align}
This tells us that if $\a_{n}=0$ for $n\geq m$, then $\g_{n\geq m} = 0$. Let us note that $\W\t$ and $v q \t$ are the small parameters (see Equation~\eqref{eq-smallparameters1}) by which we can perturbatively solve these recursion relations. Expressing the $m^{\text{th}}$ recursion relation as
\begin{align}
\sum_{n=-2}^{2}C_{m,n}\g_{m+n} &= \d_{m,0}
\end{align}
we find that $C_{m,n}$ is, in general, of the $|n|^{\text{th}}$ order in smallness. The only exception is $C_{00}$, which is of the second order in smallness because $\a_{0} = 1$ leads to an exact cancellation, a consequence of particle number conservation/gauge invariance. Because of this, to the lowest order, we find
\begin{subequations}
\begin{align}
\g_{0} &= \frac{C_{11}}{C_{00}C_{11} - C_{1,-1}C_{01}},\\
\g_{1} &= \frac{-C_{1,-1}}{C_{00}C_{11} - C_{1,-1}C_{01}}, \text{ etc.}
\end{align}
\end{subequations}
In general, all $\g_{n}$ are going to have the quantity $C_{00}C_{11} - C_{1,-1}C_{01}$ in their denominators, which, we shall show now, yields the diffusive pole for the density vertex. Indeed,
\begin{align}
&C_{00}C_{11} - C_{1,-1}C_{01}\nn\\
&\approx (- i \t \W + b_{0}v^{2}q^{2}\t^{2})(1-\a_{1}) - (-i\t v q)^{2}d_{0}f_{1}\a_{1}\nn\\
&= (1-\a_{1})\le(- i \t \W + \frac{v^{2}q^{2}\t^{2}}{3(1-\a_{1})}\ri)
\end{align}
whose zero at $\W \equiv - i D(\e) q^{2}$ yields the diffusion coefficient
\begin{align}\label{eq-generaldiffusionconstant}
D(\e) &= \frac{v^{2}\t(\e)}{3(1-\a_{1})} = \frac{v^{2}\t^{\text{tr}}(\e)}{3}
\end{align}
in terms of the transport lifetime $\t^{\text{tr}}$ (Equation~\eqref{eq-transportlifetime}), consistent with the Boltzmann approach.

To complete the current discussion, we shall write down the density vertex up to the Legendre function of order 1, which is the relevant case for Dirac/Weyl fermions:
\begin{align}\label{eq-vertexforcosine}
\G(\ve_{\mbk}, \th_{\mbk\mbq}, q, \W) &= \frac{\frac{i}{\t(\ve_{\mbk})} +  \frac{\a_{1}}{1 - \a_{1}} v q \cos\th_{\mbk\mbq}}{\W + i D(\ve_{\mbk}) q^{2}}
\end{align}

\section{Diffusion in Weyl semimetals}
\label{sec-weyldiffusion}

Semimetals have band structures where the valence and conduction bands touch at isolated point(s) in momentum space. Near these touching points the effective band theory in terms of $\mbk$, the deviation in momentum space from the touching point, is most generally given by a `Weyl' theory
\begin{align}\label{eq-weyltheory}
\mc{H}_{W} &= v \mbss\cdot \mbk
\end{align}
In general the velocities in different directions are different but we can always use appropriate anisotropic scaling transformations to obtain the `isotropic' form Equation~\eqref{eq-weyltheory}. A slowly varying background disorder potential in this theory has the form
\begin{align}\label{eq-weylpotential}
\hat{V} &= U(\mbr) \mb{1}
\end{align}
The conduction($+$)/valence($-$) bands of the Weyl theory Equation~\eqref{eq-weyltheory}, with dispersions $\ve_{\pm,\mbk}=\pm v k$, possess respectively the following wavefunctions:
\begin{subequations}
\begin{align}
\c_{\mbk, +} &= \frac{1}{\sqrt{2}}\le(\ba{c} e^{- i \f_{\mbk}}\cos\frac{\th_{\mbk}}{2}\\ \sin\frac{\th_{\mbk}}{2} \ea\ri)\\
\c_{\mbk, -} &= \frac{1}{\sqrt{2}}\le(\ba{c} e^{- i \f_{\mbk}}\sin\frac{\th_{\mbk}}{2}\\ -\cos\frac{\th_{\mbk}}{2} \ea\ri)
\end{align}
\end{subequations}
where $(\th_{\mbk}, \f_{\mbk})$ are the colatitude and azimuthal angles for $\mbk$ in polar coordinates. Using this we can show that
\begin{align}
\le|\le\la \mbk',s'| \mbk, s\ri\ra\ri|^{2} &= \frac{1 + s s' \cos\th_{\mbk\mbk'}}{2}
\end{align}
and hence, the disorder potential \eqref{eq-weylpotential} satisfying the `white noise' criterion \eqref{eq-whitenoise} has the property
\begin{align}\label{eq-weylpotentialscatter}
\frac{1}{\ms{V}}\le\la \le|U_{s\mbk,s'\mbq}\ri|^{2}\ri\ra = \frac{\z}{2}(1 + s s' \cos\th_{\mbk\mbq})
\end{align}
When $q \approx k$ and $s=s'$ in the above, it is also equal to $\mc{U}(\e_{s,\mbk},\th_{\mbk\mbq})$ by definition (Equation~\eqref{eq-Uangularavg}) and so we obtain
\begin{align}
\a_{1}=1/3, \;\a_{n>1} = 0
\end{align}
which implies, using \eqref{eq-transportlifetime},
\begin{align}\label{eq-tautrscalardisorder}
\t^{\text{tr}}(\e) &= \frac{3}{2}\t(\e)
\end{align}

\subsection{Self-Energy in the SCBA}

While our derivation of the diffusion law in Weyl semimetals will involve the self energy evaluated in the Born approximation (valid for very weak disorder $\z\to 0$), it will be instructive to evaluate it in the self-consistent Born approximation (SCBA) to expose interesting physics suggested previously\cite{1986-fradkin-yq,1986-fradkin-rt}. In the SCBA, the self energy is again given by the diagram Figure~\ref{fig-selfenergyborn}, where we interpret the electron propagator as also including the self energy correction. The self-consistent equation then becomes, with the momentum-independence of the self energy matrix $\mb{\Ss}$ leading to the simple diagonal form $\mb{\Ss} \propto \Ss \mb{1}$ ,
\begin{align}
&\Ss^{R/A}(\w,\mbk) = \z \frac{4\p}{8\p^{3}} \int_{0}^{\L} \frac{\w - \Ss^{R/A}}{(\w-\Ss^{R/A})^{2} - v^{2}p^{2}}\,p^{2}\,dp \nn\\
&= \frac{\z}{2\p^{2}v^{3}}(\w-\Ss^{R/A})\le[ - v \L + (\w-\Ss^{R/A})\tanh^{-1}\!\!\!\!\!\!\!\frac{v \L}{\w-\Ss^{R/A}} \ri]\nn\\
&\simeq \frac{\z}{2\p^{2}v^{3}}(\w-\Ss^{R/A})\le[ - v \L \mp i (\w-\Ss^{R/A})\frac{\p}{2} \ri] + \mc{O}\le(\frac{\w}{\L}\ri)
\end{align}
For $\w\ll\L$ and $\z\L < 2\p^{2}v^{2}$, the above equations can be solved to yield
\begin{align}
\Ss^{R/A}(\w,\mbk) &=\frac{\z}{2\p^{2}v^{3}}\bigg( - \frac{\w}{1-\frac{\z \L}{2\p^{2}v^{2}}} \nn\\
&\mp \frac{i \p}{2}\le(\frac{\w}{1-\frac{\z \L}{2\p^{2}v^{2}}}\ri)^{2}\bigg) + \mc{O}\le(\frac{\w}{\L}\ri)
\end{align}
This tells us that the disorder introduces a field renormalization $Z = 1-\frac{\z \L}{2\p^{2}v^{2}}$ as well as renormalizes the fermionic velocity $v \to v Z$. The latter is consistent with a QPT at $\z \approx v^{2}/\L$, when $v\to 0$ (the bandwidth collapses) and the DOS at zero energy becomes nonzero\cite{1986-fradkin-yq,1986-fradkin-rt}. Taken literally, however, since $Z\to 0+$ as one approaches this transition, the quasiparticles are destroyed both because of the loss of quasiparticle weight and a divergence of the quasiparticle damping (since $|\text{Im}(\Ss)| \propto Z^{-2}$) and it is clear that better approximation schemes are necessary to calculate the quasiparticle behavior near the transition.

In this paper, we shall be interested in the limit of weak disorder when the Born approximation is valid. Introducing a new energy scale defined using the disorder strength
\begin{align}
\mc{E} = \frac{4\p v^{3}}{\z}
\end{align}
the Born approximation is valid when
\begin{align}\label{eq-borncondition}
\mc{E}\gg \L \gg \w
\end{align}

In this approximation, the fermion propagator assumes the simple form
\begin{align}
\mb{G}^{R/A}(\w,\mbk) &= \le(\w - v \mbss\cdot\mbk \pm \frac{i}{2\t(\w)}\ri)^{-1}
\end{align}
with
\begin{align}\label{eq-weyllifetime}
\t(\w) &= \frac{2\mc{E}}{\w^{2}}
\end{align}
Upon `rotation' to the conduction/valence band basis ($s=\pm$), we recover the diagonal form \eqref{eq-greensfunctiondiagonalform} of the Green's function
\begin{align}
G^{R/A}_{s = \pm}(\w,\mbk) &= \frac{1}{\w - \ve_{s,k} \pm \frac{i}{2\t(\w)}}, \quad \ve_{s,k} = s v k
\end{align}

\subsection{The density vertex correction}

We can now use results from section~\ref{sec-anisotropicmethod} if condition~\eqref{eq-slowdos} is satisfied. Indeed, for the Weyl semimetal, the DOS is
\begin{align}\label{eq-dosweyl}
g(\w) &= \frac{\w^{2}}{2\p^{2}v^{3}}
\end{align}
and condition~\eqref{eq-slowdos} becomes equivalent to
\begin{align}
\w \ll \mc{E}
\end{align}
which is already included in the condition~\eqref{eq-borncondition}. Thus, we can use Equation~\eqref{eq-vertexforcosine} and obtain
\begin{align}\label{eq-vertexweyl}
\G(\ve_{s,\mbk}, \th_{\mbk\mbq}, q, \W) &= \frac{\frac{i}{\t(\ve_{s,\mbk})} +  \frac{v q}{2} \cos\th_{\mbk\mbq}}{\W + i D(\ve_{s,\mbk}) q^{2}}
\end{align}
The diffusion coefficient given by (using Equation~\eqref{eq-generaldiffusionconstant})
\begin{align}\label{eq-diffconstweyl}
D(\w) &= \frac{v^{2}\mc{E}}{\w^{2}}
\end{align}
and is a very sensitive function of energy, in contrast to cases where the Fermi surface is not at/close to a node in the DOS. Following the inequality~\eqref{eq-smallparameters1}, we find that the last two expressions are valid for
\begin{align}\label{eq-diffusioncondition1}
|\W|, vq \ll \frac{\w^{2}}{\mc{E}}
\end{align}

\subsection{The density response function}

\begin{figure}[h]
\begin{center}
\resizebox{5cm}{!}{\includegraphics{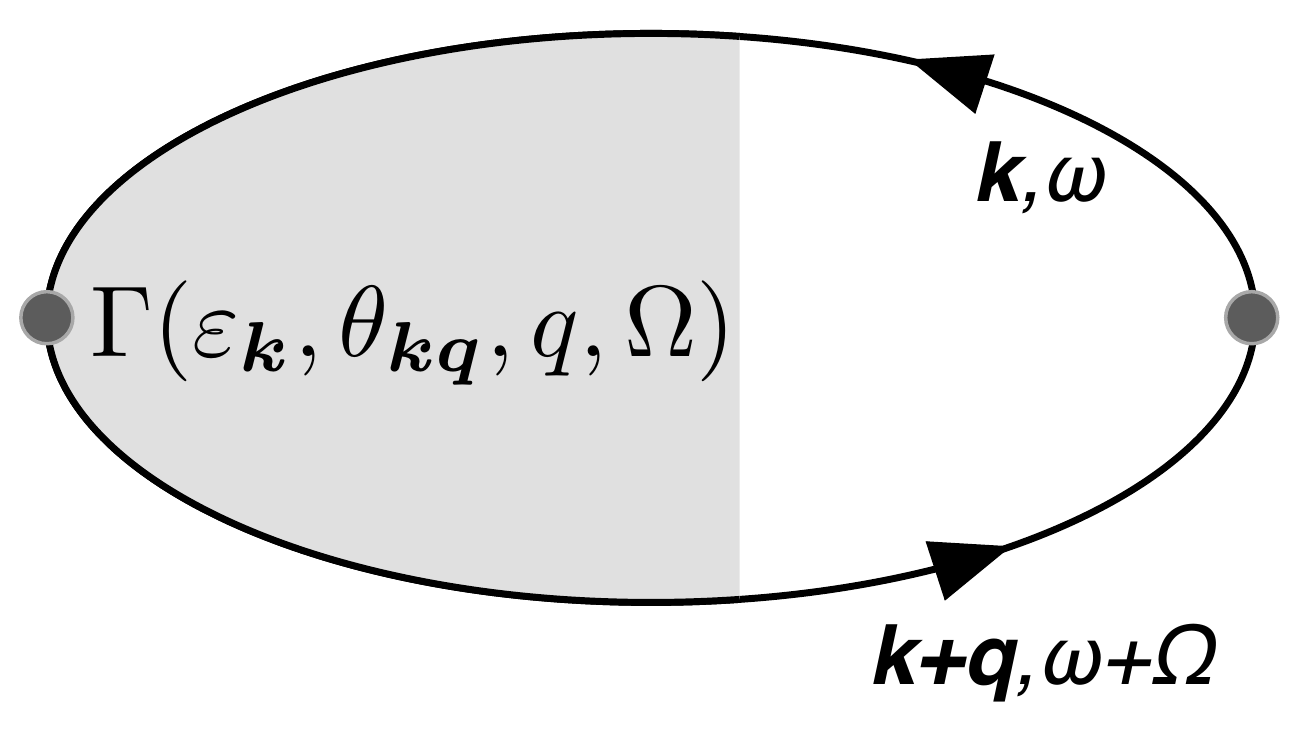}}
\caption{The charge response function in the ladder approximation. The renormalized vertex (shaded region) satisfies the Bethe Salpeter equation in Figure~\ref{fig-vertexequation}.}
\label{fig-chargeresponse}
\end{center}
\end{figure}

\begin{figure}[h]
\begin{center}
\subfigure[]{\resizebox{8cm}{!}{\includegraphics{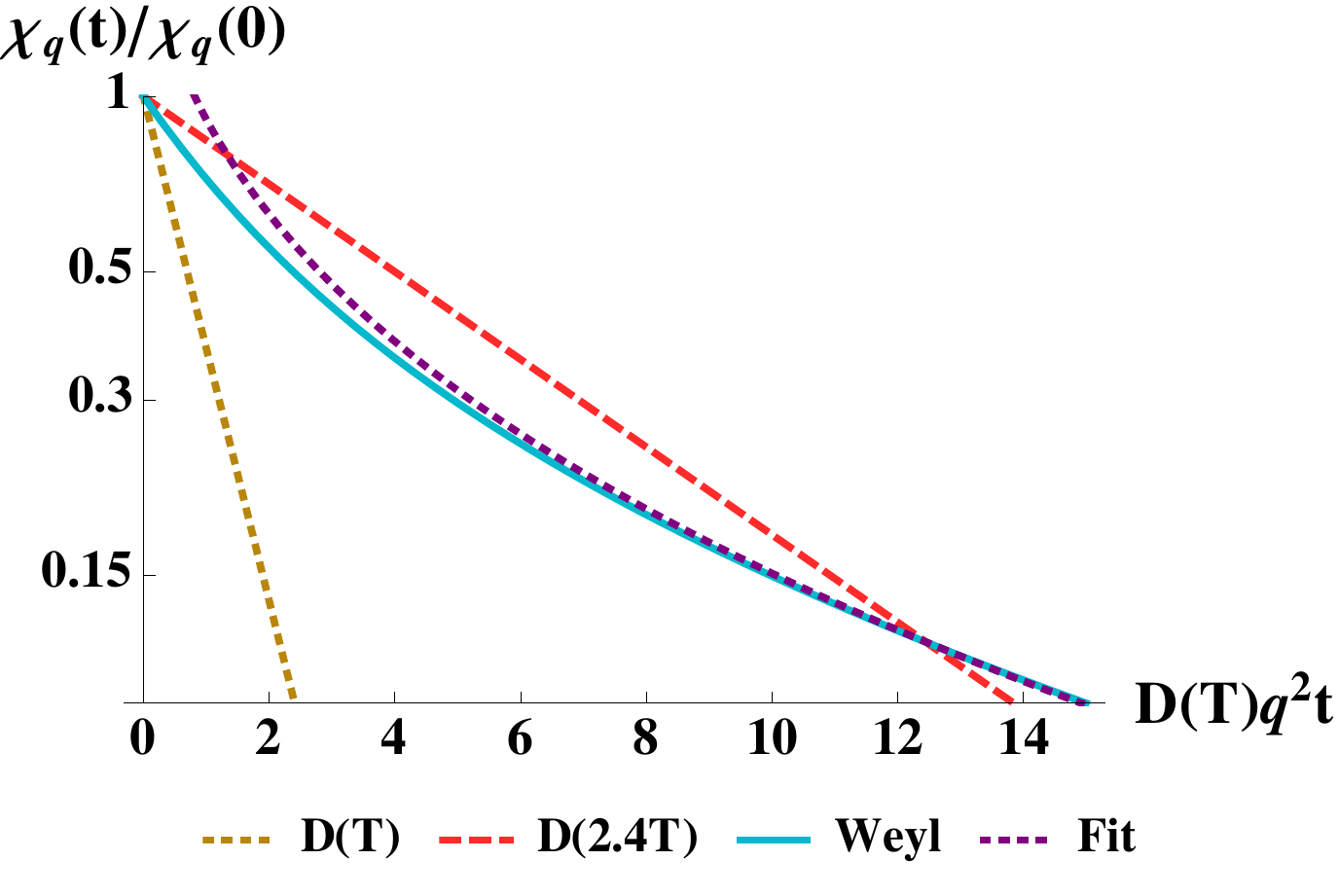}}\label{fig-chargeresponsevstimeA}}\\
\subfigure[]{\resizebox{8cm}{!}{\includegraphics{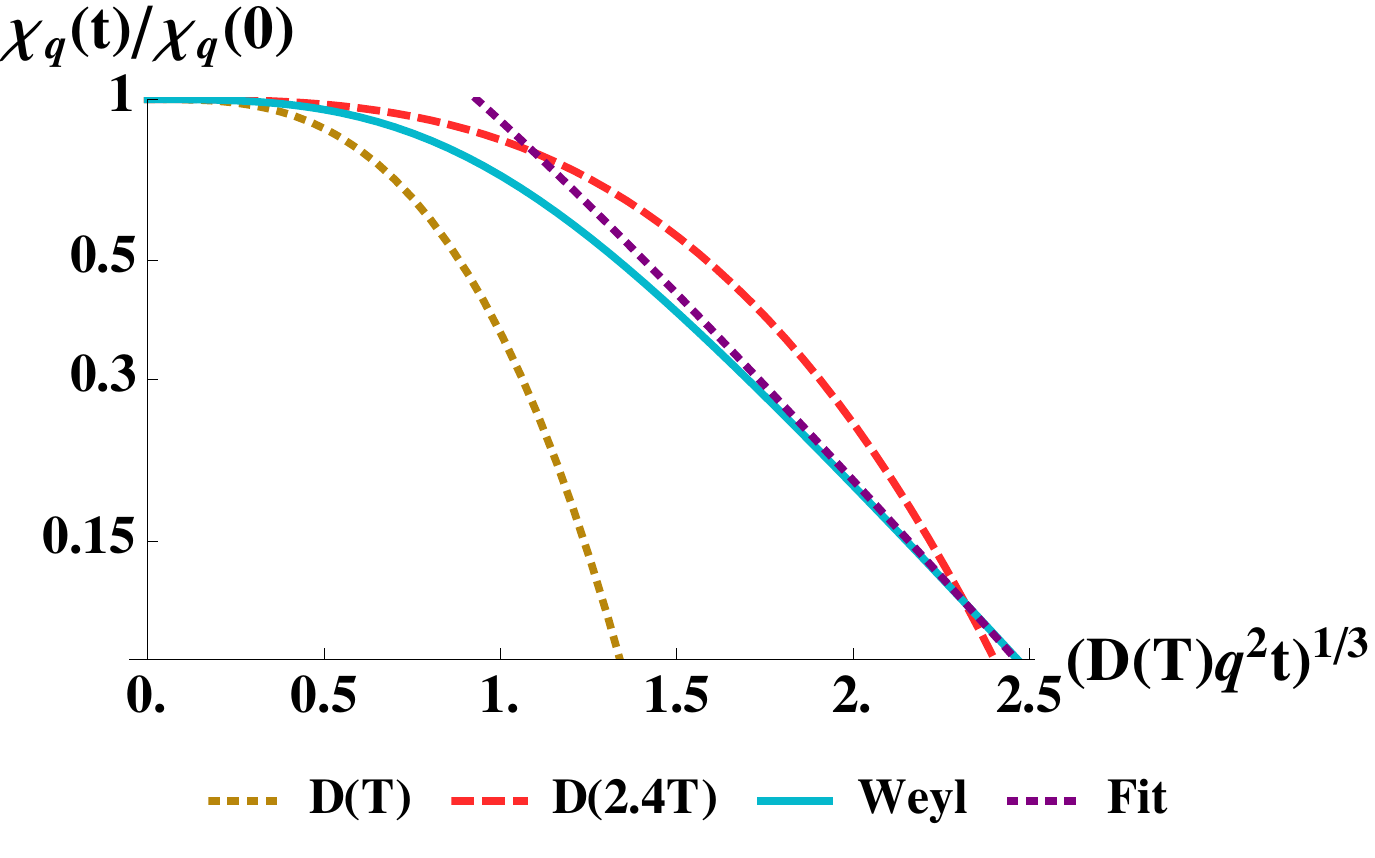}}\label{fig-chargeresponsevstimeB}}
\caption{Comparing the slow diffusive decay of $\vx_{\mbq}(t)$ in Weyl semimetals with that of particles diffusing with diffusion coefficients $D = D(T)$ or $D = D(2.4T)$, where $2.4T$ is the typical energy of the quasiparticles taking part in diffusion. The plots are log-linear so exponential decays in the quantity on the horizontal axis ($\propto t$ in \subref{fig-chargeresponsevstimeA} and $\propto t^{1/3}$ in \subref{fig-chargeresponsevstimeB}) show up as straight lines. The analytical approximation (`fit') for large times, as derived in Equation~\eqref{eq-chargeresponsetimedomain}, is also shown on these plots for comparison.}
\label{fig-chargeresponsevstime}
\end{center}
\end{figure}

The density response function \eqref{eq-densityresponsefunction1} is given by\cite{2004-bruus-lq}
\begin{align}\label{eq-chargeresponse1}
&\tilde{\vx}_{\mbq}(\W+i0) = \sum_{s=\pm}\int \frac{d\w}{2\p i}\le(n_{F}(\w+\W) - n_{F}(\w)\ri)\times\nn\\
& \int \frac{d^{3}k}{(2\p)^{3}}\G(\ve_{s,\mbk}, \th_{\mbk\mbq}, q, \W)G^{A}_{s}(\w,\mbk)G^{R}_{s}(\w+\W,\mbk+\mbq)\nn\\
&\qquad\qquad - \le(\text{value at }\mbq = 0\ri)
\end{align}
The subtraction at $\mbq=0$ is required because the bare density operator used is not `normal-ordered' while we are finding the response function for the \emph{change} in the density, which is a `normal-ordered' quantity. When condition \eqref{eq-diffusioncondition1} is satisfied for a given $\w$, using Equations~\eqref{eq-vertexweyl} and \eqref{eq-gg1eval} the product of the Green's functions above can be evaluated as follows 
\begin{align}
&\sum_{s}\int \frac{d^{3}k}{(2\p)^{3}}\G(\ve_{s,\mbk}, \th_{\mbk\mbq}, q, \W)G^{A}_{s}(\w,\mbk)G^{R}_{s}(\w+\W,\mbk+\mbq)\nn\\
&\simeq \frac{\frac{i}{\t(\w)}}{\W + i D(\w) q^{2}} \int_{-\infty}^{\infty} d\e\, \frac{g(\e)}{\le|\w - \e + \frac{i}{2\t(\w)}\ri|^{2}} \nn\\
&\simeq \frac{2 \p i g(\w)}{\W + i D(\w) q^{2}}
\end{align}
which in combination with \eqref{eq-chargeresponse1} yields,
\begin{align}\label{eq-chargeresponse2}
\tilde{\vx}_{\mbq}(\W+i0) &= \int_{-\infty}^{\infty} d\w \le(\frac{\pd n_{F}(\w)}{\pd \w}\ri) g(\w)\frac{\W}{\W + i D(\w) q^{2}}\nn\\
&\qquad\qquad - \le(\text{value at }\mbq = 0\ri)
\end{align}
\begin{figure}[h]
\begin{center}
\resizebox{7cm}{!}{\includegraphics{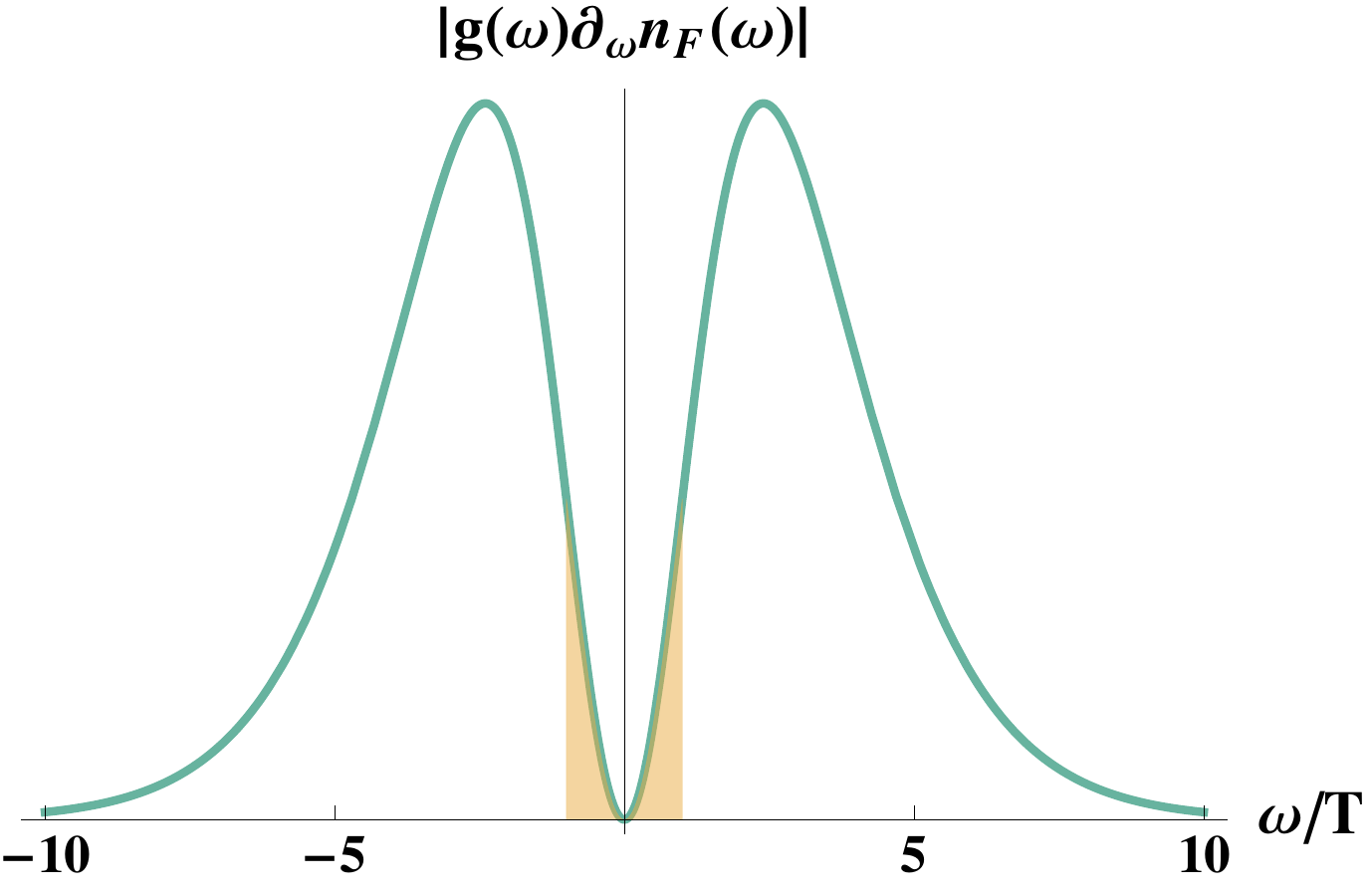}}
\caption{The contribution of particle-hole pairs at various energies to the diffusion process. The shaded region shows the negligible $\sim 4\%$ contribution from pairs with energies $|\w|<T$.}
\label{fig-diffusingenergies}
\end{center}
\end{figure}
Of course, this evaluation required the condition~\eqref{eq-diffusioncondition1} to be satisfied for all $\w$ in the above integral, which would naively require $\W \to 0$. This is because the states close to the Weyl/Dirac point have lifetimes $\t(\w\to0)\to\infty$ and hence exhibit ballistic transport. However, these states contribute negligibly to the full charge response since the weighing function $|g(\w) \pd n_{F}(\w)/\pd \w|$ in the $\w$ integral is peaked around $\w \simeq \pm 2.4 T$ and has negligible weight ($\sim 4\%$) for $|\w|<T$, as is shown in Figure~\ref{fig-diffusingenergies}. Thus the condition for observing diffusive transport is actually
\begin{align}\label{eq-diffusioncondition2}
|\W|, vq \ll \frac{1}{\t(T)} \sim \frac{T^{2}}{\mc{E}}
\end{align}
Thus, to conclude, for $\W, vq \ll T^{2}/\mc{E}\ll T$,
\begin{align}\label{eq-chargeresponse3}
\tilde{\vx}_{\mbq}(\W+i0) &= \int_{\W\t(\w)<1} \!\!\!\!\!\!\!\!d\w \le(-\frac{\pd n_{F}(\w)}{\pd \w}\ri) g(\w)\frac{i D(\w) q^{2}}{\W + i D(\w) q^{2}}
\end{align}
The lower limit in the $\w$-integral is decided by the condition for deriving the diffusive form of the density vertex, which translates to the condition $\w > \sqrt{\W\mc{E}}$. The charge response is diffusive but with a continuum of diffusive poles $D(\w) q^{2}$ ranging from $D(\w\to\infty) q^{2} = 0$ till approximately $D(T)q^{2}$. Thus, in the time domain, the charge response should be much slower than the usual exponential decay with a single timescale $(Dq^{2})^{-1}$ that is present for usual diffusion with a fixed diffusion coefficient $D$: $\vx_{\mbq}(t) \sim e^{- Dq^{2}t}$. Indeed, transforming Equation~\eqref{eq-chargeresponse3} to the time domain and using Equations~\eqref{eq-diffconstweyl} and \eqref{eq-dosweyl}, we find
\begin{align}\label{eq-chargeresponsetimedomain0}
&\vx_{\mbq}(t)\approx \Th(t) q^{2}\int_{|\w|>T} \!\!\!\! d\w \le(-\frac{\pd n_{F}(\w)}{\pd \w}\ri) g(\w)D(\w)e^{-D(\w)q^{2}t}\nn\\
&= \Th(t) \frac{\mc{E}q^{2}}{2\p^{2}v}\int_{1}^{\infty}dx \frac{e^{x}}{(e^{x} + 1)^{2}}e^{-D(T)q^{2}t/x^{2}} \quad\le(x = \frac{\e}{T}\ri)
\end{align}
This is not clear from the form of the integrand above, but this integral can be approximately calculated using the saddle point approximation\footnote{We are grateful to Tom Witten for suggesting this.}, for large times $D(T)q^{2}t\gg 1$. The saddle point of the exponent of the integrand occurs at:
\begin{align}
x_{s} &\approx (2 D(T)q^{2}t)^{1/3} + \mc{O}(e^{-(D(T)q^{2}t)^{1/3}})
\end{align}
Utilizing the usual saddle point evaluation technique, we obtain the large time behavior of the density response function:
\begin{align}\label{eq-chargeresponsetimedomain}
&\vx_{\mbq}(t) \approx \Th(t) \frac{\mc{E}q^{2}}{v\sqrt{3\p^{3}}}\le(\frac{D(T)q^{2}t}{4}\ri)^{1/6}e^{-3\le(\frac{D(T)q^{2}t}{4}\ri)^{1/3}}\nn\\
&\qquad\qquad\qquad\qquad\qquad\qquad\text{for }D(T)q^{2}t\gg1
\end{align}
Figure~\ref{fig-diffusingenergies} shows that this analytical long time expression matches the solution from \eqref{eq-chargeresponsetimedomain0} very well. This `stretched exponential' time dependence, a new physical result, is qualitatively different and slower than the usual exponential in time decay mentioned previously. This behavior is also insensitive to how the cutoff near $|\w| = T$ is handled. This is because the long time behavior arises from the contribution of quasiparticle states at higher energies that have smaller diffusion coefficients and hence slower diffusion timescales. Another point to note is that the integral expression in Equation~\eqref{eq-chargeresponsetimedomain0} is valid for timescales $t$ longer than $\t(T)$ ($\t(\w)$ is a decreasing function of $\w$)
\begin{align}
t \gg \t(T) = \frac{2\mc{E}}{T^{2}} \imply D(T)q^{2}t \gg (v q \t(T))^{2} 
\end{align}
Since $v q \t(T) \ll 1$, we can trust the answer from the aforementioned integral for almost the full range of $D(T)q^{2}t > 0$. This is plotted in Figure~\ref{fig-chargeresponsevstime}, in the form of $\vx_{\mbq}(t)/\vx_{\mbq}(0)$, and compared to the exponential decays $e^{- D q^{2}t}$ that would have resulted if all particles diffused with a diffusion coefficients $D = D(T)$ or $D = D(2.4T)$, where $2.4T$ as the typical energy of the Weyl quasiparticles taking part in diffusion (see Figure~\ref{fig-diffusingenergies} and the associated discussion).

\subsection{The diffusion memory function\\ for Weyl fermions}

We are now in a position to figure out the diffusion equation obeyed by the Weyl fermions. Comparing Equations~\eqref{eq-chidiffusionpole} and \eqref{eq-chargeresponse3}, we can solve for the memory function $\tilde{M}(z)$:
\begin{align}
\frac{- \tilde{M}_{\mbq}(\W+i0)}{\W -\tilde{M}_{\mbq}(\W+i0)} &= \frac{\tilde{\vx}_{\mbq}(\W+i0)}{\tilde{\vx}_{\mbq}(i0+)}= \int d\w \mc{P}(\w)\frac{i D(\w) q^{2}}{\W + i D(\w) q^{2}}\nn
\end{align}
where the normalized weighing function is
\begin{align}
\mc{P}(\w) &\propto \Th\le(\W\t(\w)-1\ri)\le(-\frac{\pd n_{F}(\w)}{\pd \w}\ri) g(\w) ,\int d\w\mc{P}(\w) = 1
\end{align}
The Heaviside step function $\Th$ explicitly removes the non-diffusive contribution from energies with $\W\t(\w)<1$. Manipulating the previous equations, we find
\begin{align}
\tilde{M}_{\mbq}(\W+i0) &= - i q^{2}\frac{\int d\w \frac{\mc{P}(\w) D(\w)}{\W + i D(\w) q^{2}}}{\int d\w \frac{\mc{P}(\w)}{\W + i D(\w) q^{2}}}
\end{align}
Since $D(\w)$ decreases without bound as $\w\to\infty$, there is no small timescale beyond which this memory function can take on the simple form in Equation~\eqref{eq-diffconstmemoryfunction} and so there is no (band cutoff-independent) timescale beyond which the Weyl fermions will follow a simple diffusion equation.

\begin{figure}[h]
\begin{center}
\subfigure[]{\resizebox{8cm}{!}{\includegraphics{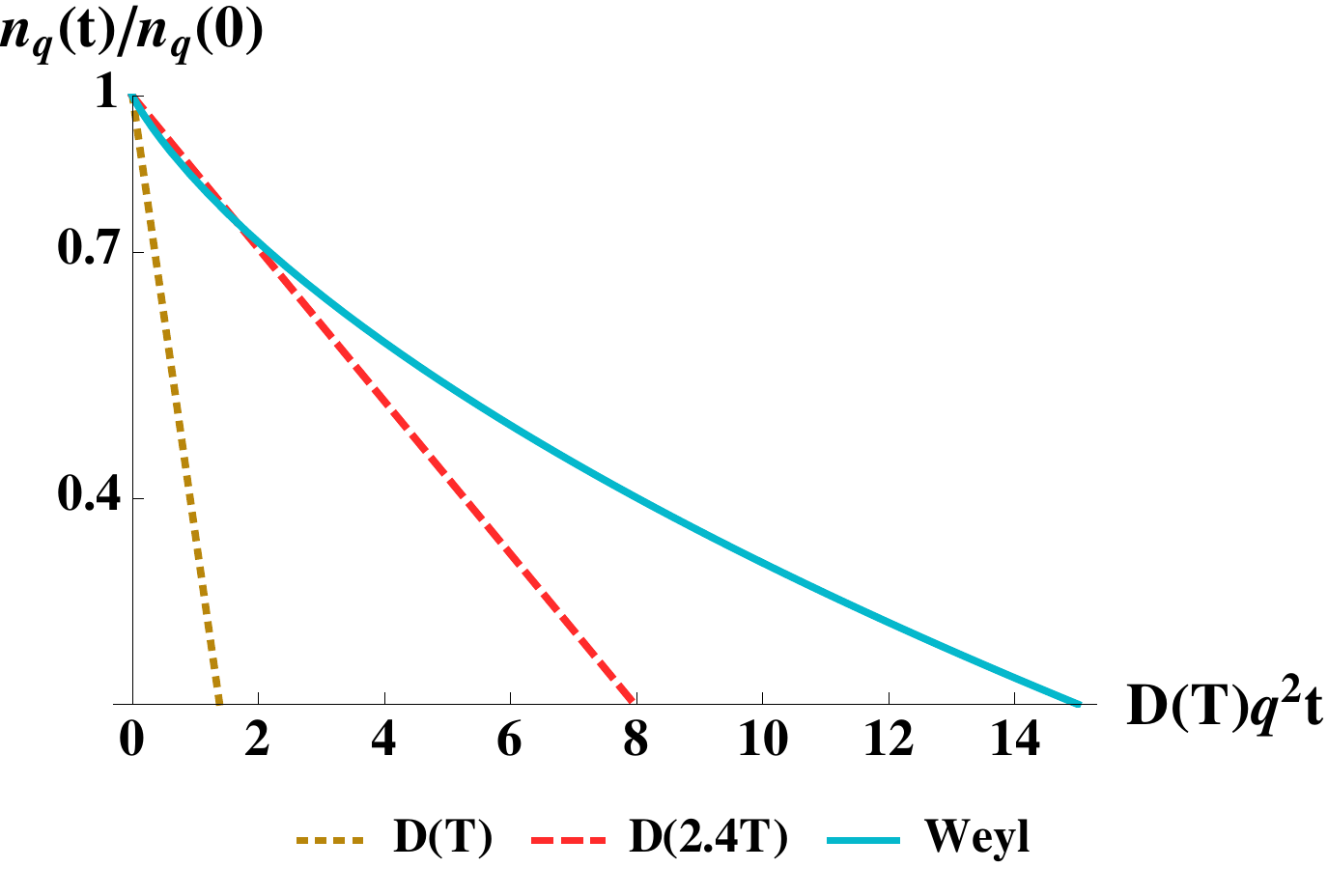}}\label{fig-chargevstimeA}}\\
\subfigure[]{\resizebox{8cm}{!}{\includegraphics{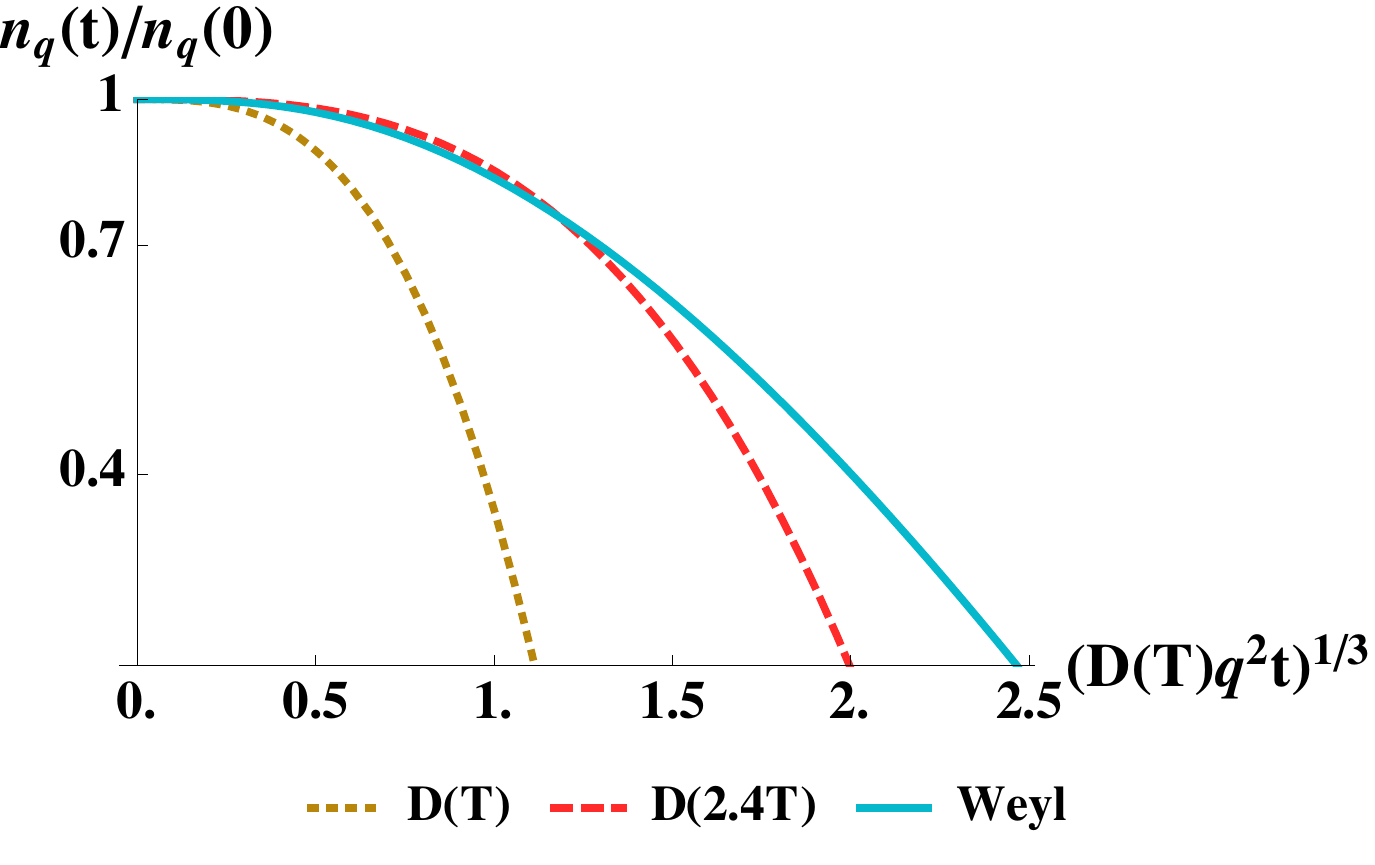}}\label{fig-chargevstimeB}}
\caption{Comparing the slow diffusive decay of a density perturbation $n_{\mbq}(t)$ in Weyl semimetals with that of conventional particles diffusing with diffusion coefficients $D = D(T)$ or $D = D(2.4T)$ according to Equation~\eqref{eq-diffusiondecay}, where $2.4T$ is the typical energy of the quasiparticles taking part in diffusion. The plots are log-linear so exponential decays in the quantity on the horizontal axis ($\propto t$ in \subref{fig-chargevstimeA} and $\propto t^{1/3}$ in \subref{fig-chargevstimeB}) show up as straight lines.}
\label{fig-chargevstime}
\end{center}
\end{figure}

This physical statement can be visualized by following the time evolution of a particle density wave $n_{\mbq}(t)$. For particles that satisfy a diffusion equation with diffusion coefficient $D$, Equation~\eqref{eq-diffusionequationmaster} tells us that $n_{\mbq}(t)/n_{\mbq}(0) \sim e^{- D q^{2}t}$. However, for the Weyl case, using Equations~\eqref{eq-diffusionmemorydef1} and \eqref{eq-chidiffusionpole}, we find
\begin{align}
\frac{\tilde{n}_{\mbq}(\W+i0)}{n_{\mbq}(0)} &= \frac{1}{\W}\le(1 - \frac{\tilde{\vx}_{\mbq}(\W+i0)}{\tilde{\vx}_{\mbq}(i0+)}\ri)\nn\\
&\propto \int d\w \le(-\frac{\pd n_{F}(\w)}{\pd \w}\ri) \frac{g(\w)}{\W + i D(\w) q^{2}}
\end{align}
This means that for long times $t\gg\t(T)$
\begin{align}\label{eq-chargeresponse4}
&\frac{n_{\mbq}(t)}{n_{\mbq}(0)} \propto  \int d\w \le(-\frac{\pd n_{F}(\w)}{\pd \w}\ri) g(\w) e^{-D(\w)q^{2}t}\nn\\
&\propto \int_{0}^{\infty} dx\, x^{2} \frac{e^{x}}{(e^{x} + 1)^{2}}e^{-D(T)q^{2}t/x^{2}} \quad\le(x = \frac{\e}{T}\ri)\nn\\
&\approx 4\sqrt{\frac{\p}{3}} \le(\frac{D(T)q^{2}t}{4}\ri)^{1/6} e^{-3\le(\frac{D(T)q^{2}t}{4}\ri)^{1/3}}\nn\\
&\qquad\qquad\qquad\qquad\qquad\qquad \text{ for }D(T)q^{2}t\gg 1
\end{align}
This stretched exponential decay, found using a saddle point analysis, is similar to that for the density response function (Figure~\ref{fig-chargeresponsevstime}). The numerically evaluated integral is compared against the exponential decays~\eqref{eq-diffusiondecay} of a density perturbation for conventional particles diffusing with a diffusion coefficients $D(T)$ or $D(2.4T)$ in Figure~\ref{fig-chargevstime}.\footnote{The convergence of the saddle point estimate to the actual value of the integral is very slow as $t\to\infty$ and has not been included in Figure~\ref{fig-chargevstime}.}

\subsection{A phenomenological picture of Weyl diffusion}

Equations~\eqref{eq-chargeresponse3} will now be derived from a simple phenomenological model of Weyl fermion quasiparticles, with particles at different energies diffusing independently with an energy-dependent diffusion coefficient $D(\e)$. This clarifies the physical picture of the unconventional behavior uncovered in the past few subsections. Denoting the density of particles at energy $\e$ by the quantity $n_{\mbq}(t|\e)$, we should have
\begin{align}\label{eq-totaldensity}
n_{\mbq}(t) &= \int d\e \;n_{\mbq}(t|\e)
\end{align}
At $t=0$, we shall assume thermal equilibrium in the presence of a frozen fluctuation $\m_{\mbq}$ in the chemical potential. In that case
\begin{align}\label{eq-initialthermalized}
n_{\mbq}(t = 0) &= \int d\e\; g(\e)\le(-\frac{\pd n_{F}(\e)}{\pd \e}\ri)\m_{\mbq}\nn\\
\imply n_{\mbq}(t=0|\e) &= g(\e)\le(-\frac{\pd n_{F}(\e)}{\pd \e}\ri)\m_{\mbq}
\end{align}
and also
\begin{align}\label{eq-phenomstaticchi}
\tilde{\vx}_{\mbq}(i0+) &= \frac{n_{\mbq}(t = 0)}{\m_{\mbq}} = \int d\e\; g(\e)\le(-\frac{\pd n_{F}(\e)}{\pd \e}\ri)
\end{align}
Following the discussion surrounding Figure~\ref{fig-diffusingenergies}, we see that $n_{\mbq}(t=0|\e)$ is negligible for $\e<T$ and thus we should disregard them in the discussion below. Now, we shall relax the chemical potential to the constant equilibrium value and calculate the diffusive process by which the numbers relax to their equilibrium (zero) values. Using Equation~\eqref{eq-diffusionequationmaster},
\begin{align}\label{eq-diffusionmaster2}
\tilde{n}_{\mbq}(z|\e) &= \frac{n_{\mbq}(t=0|\e)}{z + i D(\e) q^{2}},\quad |z|\t(\e) \ll 1\nn\\
&= \le(-\frac{\pd n_{F}(\e)}{\pd \e}\ri)\frac{g(\e)}{z + i D(\e) q^{2}}\m_{\mbq}
\end{align}
Thus, using Equation~\eqref{eq-totaldensity}, the total density response is
\begin{align}
\tilde{n}_{\mbq}(z) &= \int_{|z|\t(\e)<1} d\e \;\le(-\frac{\pd n_{F}(\e)}{\pd \e}\ri)\frac{g(\e)}{z + i D(\e) q^{2}}\m_{\mbq}\nn\\
&\equiv \frac{n_{\mbq}(0)}{z - \tilde{M}_{\mbq}(z)} \quad (\text{using \eqref{eq-diffusionmemorydef1}})
\end{align}
and so
\begin{align}
\frac{\tilde{\vx}_{\mbq}(i0+)}{z - \tilde{M}_{\mbq}(z)} &= \int_{|z|\t(\e)<1} \!\!\!\!\!\!\!\!d\e \;\le(-\frac{\pd n_{F}(\e)}{\pd \e}\ri)\frac{g(\e)}{z + i D(\e) q^{2}}
\end{align}
Finally, using Equations~\eqref{eq-chidiffusionpole} and \eqref{eq-phenomstaticchi},
\begin{align}\label{eq-chidiffusionpole2}
\tilde{\vx}_{\mbq}(z) &= \le(1-\frac{z}{z -\tilde{M}_{\mbq}(z)}\ri)\tilde{\vx}_{\mbq}(i0+)\nn\\
&= \int_{\W\t(\w)<1} d\e \;\le(-\frac{\pd n_{F}(\e)}{\pd \e}\ri)g(\e)\frac{ i D(\e) q^{2}}{z + i D(\e) q^{2}}
\end{align}
which is identical to Equation~\eqref{eq-chargeresponse3} and proves the applicability of this simple picture to the previous results of this section.

\section{Diffusive conductivity, vector disorder and Coulomb interactions}
\label{sec-leftover}

We conclude the main body of this paper with comments about the conductivity of disordered Weyl semimetals at finite temperature and the effects of including isotropic `vector' disorder, i.e, disorder terms proportional to the Pauli matrices in \eqref{eq-weyltheory}, assuming rotational isotropy.

\subsection{Conductivity}

Using the continuity equation $-e\pd_{t}\rr + \grad\cdot\mbJ = 0$, the (longitudinal) conductivity $\s = \s_{xx} = \s_{yy} = \s_{zz}$ may be related to the density response function $\vx$:
\begin{align}
\s_{\mbq}(\W) &= - i e^{2} \frac{\W}{q^{2}}\vx_{\mbq}(\W)
\end{align}
In the diffusing limit $\W, vq \ll T^{2}/\mc{E}\ll T$, we can use \eqref{eq-chargeresponse3} to obtain
\begin{align}\label{eq-conductivityweyl}
\s_{\mbq}(\W) &\simeq \int_{\W\t(\w)<1} \!\!\!\!\!\!d\w \le(-\frac{\pd n_{F}(\w)}{\pd \w}\ri) g(\w)\frac{e^{2}\W D(\w)}{\W + i D(\w) q^{2}}
\end{align}
Taking the limit $q\to 0$ and then $\W\to0$, we arrive at the DC conductivity
\begin{align}
\s_{\text{DC}} &\simeq e^{2} \int d\w \le(-\frac{\pd n_{F}(\w)}{\pd \w}\ri) g(\w)D(\w) \nn\\
&= \frac{e^{2}\mc{E}}{2\p^{2}v} \equiv \frac{4 e^{2} v^{2}}{\z h}
\end{align}
where the Planck's constant $h$ has been restored and the result is re-expressed in terms of $\z$ defined in \eqref{eq-whitenoise}. This answer is equivalent to earlier results\cite{2012-hosur-uq, 2011-burkov-uq}, but with the correct numerical factors\cite{2012-hosur-uq} and the renormalization of the current vertex\cite{2004-bruus-lq}, equal to the ratio $\t^{\text{tr}}/\t = 3/2$.

In addition, for small $\W$ the diffusive component of the conductivity decreases from its DC value by a term proportional to $\sqrt{\W}$, as has been obtained previously\cite{2011-burkov-uq} (but without using the conserving approximation or taking into account the diffusive-ballistic crossover). This arises due to the $\sim\sqrt{\W\mc{E}}$ behavior of the lower limit of the $\w$ integral in \eqref{eq-conductivityweyl}, below which ballistic transport occurs. However, in this energy range the calculations need to incorporate the diffusive-ballistic transition accurately, as well as inter-band scattering processes in order to correctly obtain the coefficient of this $\sqrt{\W}$ term.

\subsection{Vector disorder}

In addition to the usual `potential' disorder term $U(\mbr)\mb{1}$, we can also include `vector' disorder in the Weyl theory~\eqref{eq-weyltheory}
\begin{align}
\mc{H}_{V} &= \mbs{V}(\mbr)\cdot\mbss
\end{align}
We assume `white' noise disorder with rotational symmetry restored on the average, which means that
\begin{align}
\le\la V_{a}(\mbr)V_{b}(\mbr')\ri\ra &= \d_{ab} \,\z_{V}\d(\mbr-\mbr') 
\end{align}
and also that this `vector' disorder is uncorrelated with the `potential' disorder on the average $\le\la V_{a} U\ri\ra = 0$. With these assumptions, we can show that the net effect is to change the effective interaction induced by disorder, Equation~\eqref{eq-weylpotentialscatter}, to the form
\begin{align}\label{eq-weylpotentialscatter}
\frac{\le\la \le|U_{\mbk\mbq}^{\text{total}}\ri|^{2}\ri\ra}{\ms{V}} = \frac{\z}{2}(1 + s s' \cos\th_{\mbk\mbq}) + \frac{3\z_{V}}{2}\le(1 - \frac{s s'}{3} \cos\th_{\mbk\mbq}\ri)
\end{align}
This means that depending on the relative strengths of the two kinds of disorder, we have
\begin{align}
-\frac{1}{9} \leq \a_{1} \leq \frac{1}{3}
\end{align}
and so
\begin{align}\label{eq-tautrvectordisorder}
\frac{9}{10} \leq \frac{\t^{\text{tr}}}{\t} \leq \frac{3}{2}
\end{align}
Apart from this modification the rest of the physics is the same as when only `potential' disorder is present.

\subsection{Coulomb interactions}

Unlike short range quenched disorder which is an irrelevant perturbation to the Weyl theory, long range instantaneous Coulomb interaction is a marginal perturbation and results in strong electron-electron scattering (including inelastic processes) as the temperature is decreased. This phenomenon can be approximately treated using the Quantum Boltzmann Equation and shows that the conductivity goes to zero as $T\to 0$\cite{2008-fritz-yq,2012-hosur-uq}. Thus, in the presence of Coulomb interactions, our results for quenched disorder limited transport hold only for high enough temperatures, when the Coulomb scattering may be neglected in comparison to that due to quenched disorder.

\section{Discussion}

\begin{figure}[ht]
\begin{center}
\resizebox{9cm}{!}{\includegraphics[trim=3cm 0cm 3cm 0cm, clip=true, angle=0]{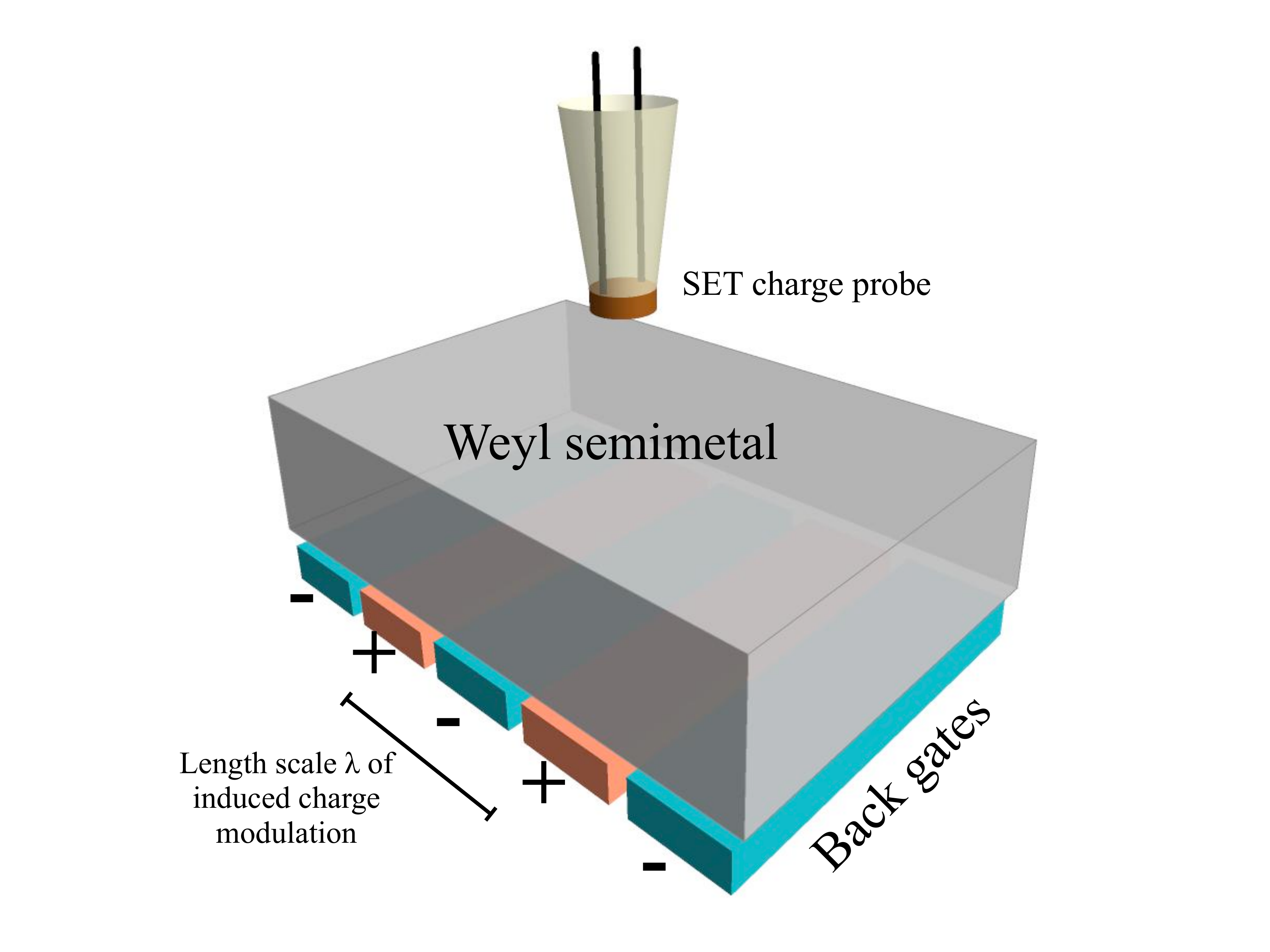}}
\caption{Experimental setup for measuring charge diffusion in a Weyl semimetal. The back gate arrangement induces a charge density modulation in the Weyl semimetal with a characteristic wavelength $\l = 2\p/q$ that is equal to the distance between two successive gates with the same polarity. At $t=0$, the gates are switched off and the charge modulation relaxes to the equilibrium uniform value and this decay is tracked using, say, an SET-based charge-sensitive probe\cite{2007-martin-rt}. We predict that the charge decay will follow the form predicted in Figure~\ref{fig-chargevstime}.}
\label{fig-exptsetup}
\end{center}
\end{figure}

In this paper, we have initially introduced the formalism required to compute correctly the renormalized charge vertex, in the diffuson approximation, in the presence of disorder with an anisotropic scattering amplitude. We have then used this to study the case of Weyl fermions in the presence of quenched disorder in the quantum critical regime when the chemical potential is small compared to the temperature. These quasiparticles exhibit anisotropic scattering and their DOS varies sharply with energy near the Fermi point. The anisotropic scattering results in the transport timescale being the relevant quantity in the diffusion parameter, the quantum critical nature of the problem results in temperature being the only low energy scale in the theory and the sharp linear increase of the DOS leads to a very slow unconventional diffusion process characterized by a memory function which does not have a sharp upper limit on the timescales involved.

When a Weyl semimetal is found in the laboratory, this novel slow diffusive relaxation may be observed as a test of our theory. The experimental setup would involve preparing a sample of the Weyl semimetal with a frozen long-wavelength excess charge modulation at $t=0$ induced by, say, a backgate arrangement that is electrically insulated from the sample, as shown schematically in Figure~\ref{fig-exptsetup}. The backgate is then grounded and the induced charge is allowed to relax to the steady state uniform distribution. The amount of charge present at a given time may be sensed using, say, a sensitive SET charge probe\cite{2007-martin-rt}. The relaxation of charge is then predicted to follow the slow process shown in Figure~\ref{fig-chargevstime}, with the value of $q$ being provided by the inverse wavelength of the induced charge modulation, which may be tuned by changing the separation between the backgates. The initial decay may, however, turn out to be much faster in the experiment, due to charge diffusion along the surface Fermi `arcs'.

\begin{acknowledgments}
This research was supported by the Institute of Condensed Matter Theory at UIUC. We would like to acknowledge stimulating discussions with Eduardo Fradkin, Ilya Gruzberg, Leo Kadanoff and Tom Witten.
\end{acknowledgments}


\begin{thebibliography}{36}%
\makeatletter
\providecommand \@ifxundefined [1]{%
 \@ifx{#1\undefined}
}%
\providecommand \@ifnum [1]{%
 \ifnum #1\expandafter \@firstoftwo
 \else \expandafter \@secondoftwo
 \fi
}%
\providecommand \@ifx [1]{%
 \ifx #1\expandafter \@firstoftwo
 \else \expandafter \@secondoftwo
 \fi
}%
\providecommand \natexlab [1]{#1}%
\providecommand \enquote  [1]{``#1''}%
\providecommand \bibnamefont  [1]{#1}%
\providecommand \bibfnamefont [1]{#1}%
\providecommand \citenamefont [1]{#1}%
\providecommand \href@noop [0]{\@secondoftwo}%
\providecommand \href [0]{\begingroup \@sanitize@url \@href}%
\providecommand \@href[1]{\@@startlink{#1}\@@href}%
\providecommand \@@href[1]{\endgroup#1\@@endlink}%
\providecommand \@sanitize@url [0]{\catcode `\\12\catcode `\$12\catcode
  `\&12\catcode `\#12\catcode `\^12\catcode `\_12\catcode `\%12\relax}%
\providecommand \@@startlink[1]{}%
\providecommand \@@endlink[0]{}%
\providecommand \url  [0]{\begingroup\@sanitize@url \@url }%
\providecommand \@url [1]{\endgroup\@href {#1}{\urlprefix }}%
\providecommand \urlprefix  [0]{URL }%
\providecommand \Eprint [0]{\href }%
\providecommand \doibase [0]{http://dx.doi.org/}%
\providecommand \selectlanguage [0]{\@gobble}%
\providecommand \bibinfo  [0]{\@secondoftwo}%
\providecommand \bibfield  [0]{\@secondoftwo}%
\providecommand \translation [1]{[#1]}%
\providecommand \BibitemOpen [0]{}%
\providecommand \bibitemStop [0]{}%
\providecommand \bibitemNoStop [0]{.\EOS\space}%
\providecommand \EOS [0]{\spacefactor3000\relax}%
\providecommand \BibitemShut  [1]{\csname bibitem#1\endcsname}%
\let\auto@bib@innerbib\@empty
\bibitem [{\citenamefont {Geim}(2011)}]{2011-geim-fk}%
  \BibitemOpen
  \bibfield  {author} {\bibinfo {author} {\bibfnamefont {A.~K.}\ \bibnamefont
  {Geim}},\ }\href {\doibase 10.1103/RevModPhys.83.851} {\bibfield  {journal}
  {\bibinfo  {journal} {Rev. Mod. Phys.}\ }\textbf {\bibinfo {volume} {83}},\
  \bibinfo {pages} {851} (\bibinfo {year} {2011})}\BibitemShut {NoStop}%
\bibitem [{\citenamefont {Hasan}\ and\ \citenamefont
  {Moore}(2011)}]{2011-hasan-uq}%
  \BibitemOpen
  \bibfield  {author} {\bibinfo {author} {\bibfnamefont {M.~Z.}\ \bibnamefont
  {Hasan}}\ and\ \bibinfo {author} {\bibfnamefont {J.~E.}\ \bibnamefont
  {Moore}},\ }\href {\doibase 10.1146/annurev-conmatphys-062910-140432}
  {\bibfield  {journal} {\bibinfo  {journal} {Annu. Rev. Cond. Mat. Phys.}\
  }\textbf {\bibinfo {volume} {2}},\ \bibinfo {pages} {55} (\bibinfo {year}
  {2011})}\BibitemShut {NoStop}%
\bibitem [{\citenamefont {Nielsen}\ and\ \citenamefont
  {Ninomiya}(1981)}]{1981-nielsen-mz}%
  \BibitemOpen
  \bibfield  {author} {\bibinfo {author} {\bibfnamefont {H.~B.}\ \bibnamefont
  {Nielsen}}\ and\ \bibinfo {author} {\bibfnamefont {M.}~\bibnamefont
  {Ninomiya}},\ }\href {\doibase DOI: 10.1016/0550-3213(81)90361-8} {\bibfield
  {journal} {\bibinfo  {journal} {Nuclear Physics B}\ }\textbf {\bibinfo
  {volume} {185}},\ \bibinfo {pages} {20 } (\bibinfo {year}
  {1981})}\BibitemShut {NoStop}%
\bibitem [{\citenamefont {Bernevig}\ and\ \citenamefont
  {Hughes}(2013)}]{2013-bernevig-qy}%
  \BibitemOpen
  \bibfield  {author} {\bibinfo {author} {\bibfnamefont {B.~A.}\ \bibnamefont
  {Bernevig}}\ and\ \bibinfo {author} {\bibfnamefont {T.~L.}\ \bibnamefont
  {Hughes}},\ }\href {http://books.google.com/books?id=wOn7JHSSxrsC} {\emph
  {\bibinfo {title} {Topological Insulators and Topological Superconductors}}}\
  (\bibinfo  {publisher} {Princeton University Press},\ \bibinfo {year}
  {2013})\BibitemShut {NoStop}%
\bibitem [{\citenamefont {Drut}\ and\ \citenamefont
  {L\"ahde}(2009)}]{2009-drut-uq}%
  \BibitemOpen
  \bibfield  {author} {\bibinfo {author} {\bibfnamefont {J.~E.}\ \bibnamefont
  {Drut}}\ and\ \bibinfo {author} {\bibfnamefont {T.~A.}\ \bibnamefont
  {L\"ahde}},\ }\href {\doibase 10.1103/PhysRevLett.102.026802} {\bibfield
  {journal} {\bibinfo  {journal} {Phys. Rev. Lett.}\ }\textbf {\bibinfo
  {volume} {102}},\ \bibinfo {pages} {026802} (\bibinfo {year}
  {2009})}\BibitemShut {NoStop}%
\bibitem [{\citenamefont {Biswas}\ and\ \citenamefont
  {Balatsky}(2010)}]{2010-biswas-uq}%
  \BibitemOpen
  \bibfield  {author} {\bibinfo {author} {\bibfnamefont {R.~R.}\ \bibnamefont
  {Biswas}}\ and\ \bibinfo {author} {\bibfnamefont {A.~V.}\ \bibnamefont
  {Balatsky}},\ }\href {\doibase 10.1103/PhysRevB.81.233405} {\bibfield
  {journal} {\bibinfo  {journal} {Phys. Rev. B}\ }\textbf {\bibinfo {volume}
  {81}},\ \bibinfo {pages} {233405} (\bibinfo {year} {2010})}\BibitemShut
  {NoStop}%
\bibitem [{\citenamefont {Wray}\ \emph {et~al.}(2011)\citenamefont {Wray},
  \citenamefont {Xu}, \citenamefont {Xia}, \citenamefont {Hsieh}, \citenamefont
  {Fedorov}, \citenamefont {Hor}, \citenamefont {Cava}, \citenamefont {Bansil},
  \citenamefont {Lin},\ and\ \citenamefont {Hasan}}]{2011-wray-fk}%
  \BibitemOpen
  \bibfield  {author} {\bibinfo {author} {\bibfnamefont {L.~A.}\ \bibnamefont
  {Wray}}, \bibinfo {author} {\bibfnamefont {S.-Y.}\ \bibnamefont {Xu}},
  \bibinfo {author} {\bibfnamefont {Y.}~\bibnamefont {Xia}}, \bibinfo {author}
  {\bibfnamefont {D.}~\bibnamefont {Hsieh}}, \bibinfo {author} {\bibfnamefont
  {A.~V.}\ \bibnamefont {Fedorov}}, \bibinfo {author} {\bibfnamefont {Y.~S.}\
  \bibnamefont {Hor}}, \bibinfo {author} {\bibfnamefont {R.~J.}\ \bibnamefont
  {Cava}}, \bibinfo {author} {\bibfnamefont {A.}~\bibnamefont {Bansil}},
  \bibinfo {author} {\bibfnamefont {H.}~\bibnamefont {Lin}}, \ and\ \bibinfo
  {author} {\bibfnamefont {M.~Z.}\ \bibnamefont {Hasan}},\ }\href {\doibase
  10.1038/nphys1838} {\bibfield  {journal} {\bibinfo  {journal} {Nat Phys}\
  }\textbf {\bibinfo {volume} {7}},\ \bibinfo {pages} {32} (\bibinfo {year}
  {2011})}\BibitemShut {NoStop}%
\bibitem [{\citenamefont {Wan}\ \emph {et~al.}(2011)\citenamefont {Wan},
  \citenamefont {Turner}, \citenamefont {Vishwanath},\ and\ \citenamefont
  {Savrasov}}]{2011-wan-fk}%
  \BibitemOpen
  \bibfield  {author} {\bibinfo {author} {\bibfnamefont {X.}~\bibnamefont
  {Wan}}, \bibinfo {author} {\bibfnamefont {A.~M.}\ \bibnamefont {Turner}},
  \bibinfo {author} {\bibfnamefont {A.}~\bibnamefont {Vishwanath}}, \ and\
  \bibinfo {author} {\bibfnamefont {S.~Y.}\ \bibnamefont {Savrasov}},\ }\href
  {\doibase 10.1103/PhysRevB.83.205101} {\bibfield  {journal} {\bibinfo
  {journal} {Phys. Rev. B}\ }\textbf {\bibinfo {volume} {83}},\ \bibinfo
  {pages} {205101} (\bibinfo {year} {2011})}\BibitemShut {NoStop}%
\bibitem [{\citenamefont {Burkov}\ and\ \citenamefont
  {Balents}(2011)}]{2011-burkov-fk}%
  \BibitemOpen
  \bibfield  {author} {\bibinfo {author} {\bibfnamefont {A.~A.}\ \bibnamefont
  {Burkov}}\ and\ \bibinfo {author} {\bibfnamefont {L.}~\bibnamefont
  {Balents}},\ }\href {\doibase 10.1103/PhysRevLett.107.127205} {\bibfield
  {journal} {\bibinfo  {journal} {Phys. Rev. Lett.}\ }\textbf {\bibinfo
  {volume} {107}},\ \bibinfo {pages} {127205} (\bibinfo {year}
  {2011})}\BibitemShut {NoStop}%
\bibitem [{\citenamefont {Balents}(2011)}]{2011-balents-fj}%
  \BibitemOpen
  \bibfield  {author} {\bibinfo {author} {\bibfnamefont {L.}~\bibnamefont
  {Balents}},\ }\href {\doibase 10.1103/Physics.4.36} {\bibfield  {journal}
  {\bibinfo  {journal} {Physics}\ }\textbf {\bibinfo {volume} {4}},\ \bibinfo
  {pages} {36} (\bibinfo {year} {2011})}\BibitemShut {NoStop}%
\bibitem [{\citenamefont {Son}\ and\ \citenamefont
  {Yamamoto}(2012)}]{2012-son-yq}%
  \BibitemOpen
  \bibfield  {author} {\bibinfo {author} {\bibfnamefont {D.~T.}\ \bibnamefont
  {Son}}\ and\ \bibinfo {author} {\bibfnamefont {N.}~\bibnamefont {Yamamoto}},\
  }\href {\doibase 10.1103/PhysRevLett.109.181602} {\bibfield  {journal}
  {\bibinfo  {journal} {Phys. Rev. Lett.}\ }\textbf {\bibinfo {volume} {109}},\
  \bibinfo {pages} {181602} (\bibinfo {year} {2012})}\BibitemShut {NoStop}%
\bibitem [{\citenamefont {Fritz}\ \emph {et~al.}(2008)\citenamefont {Fritz},
  \citenamefont {Schmalian}, \citenamefont {M\"{u}ller},\ and\ \citenamefont
  {Sachdev}}]{2008-fritz-yq}%
  \BibitemOpen
  \bibfield  {author} {\bibinfo {author} {\bibfnamefont {L.}~\bibnamefont
  {Fritz}}, \bibinfo {author} {\bibfnamefont {J.}~\bibnamefont {Schmalian}},
  \bibinfo {author} {\bibfnamefont {M.}~\bibnamefont {M\"{u}ller}}, \ and\
  \bibinfo {author} {\bibfnamefont {S.}~\bibnamefont {Sachdev}},\ }\href
  {\doibase 10.1103/PhysRevB.78.085416} {\bibfield  {journal} {\bibinfo
  {journal} {Phys. Rev. B}\ }\textbf {\bibinfo {volume} {78}},\ \bibinfo {eid}
  {085416} (\bibinfo {year} {2008})}\BibitemShut {NoStop}%
\bibitem [{\citenamefont {Hosur}\ \emph {et~al.}(2012)\citenamefont {Hosur},
  \citenamefont {Parameswaran},\ and\ \citenamefont
  {Vishwanath}}]{2012-hosur-uq}%
  \BibitemOpen
  \bibfield  {author} {\bibinfo {author} {\bibfnamefont {P.}~\bibnamefont
  {Hosur}}, \bibinfo {author} {\bibfnamefont {S.~A.}\ \bibnamefont
  {Parameswaran}}, \ and\ \bibinfo {author} {\bibfnamefont {A.}~\bibnamefont
  {Vishwanath}},\ }\href {\doibase 10.1103/PhysRevLett.108.046602} {\bibfield
  {journal} {\bibinfo  {journal} {Phys. Rev. Lett.}\ }\textbf {\bibinfo
  {volume} {108}},\ \bibinfo {pages} {046602} (\bibinfo {year}
  {2012})}\BibitemShut {NoStop}%
\bibitem [{\citenamefont {Burkov}\ \emph {et~al.}(2011)\citenamefont {Burkov},
  \citenamefont {Hook},\ and\ \citenamefont {Balents}}]{2011-burkov-uq}%
  \BibitemOpen
  \bibfield  {author} {\bibinfo {author} {\bibfnamefont {A.~A.}\ \bibnamefont
  {Burkov}}, \bibinfo {author} {\bibfnamefont {M.~D.}\ \bibnamefont {Hook}}, \
  and\ \bibinfo {author} {\bibfnamefont {L.}~\bibnamefont {Balents}},\ }\href
  {\doibase 10.1103/PhysRevB.84.235126} {\bibfield  {journal} {\bibinfo
  {journal} {Phys. Rev. B}\ }\textbf {\bibinfo {volume} {84}},\ \bibinfo
  {pages} {235126} (\bibinfo {year} {2011})}\BibitemShut {NoStop}%
\bibitem [{\citenamefont {Sheehy}\ and\ \citenamefont
  {Schmalian}(2007)}]{2007-sheehy-sv}%
  \BibitemOpen
  \bibfield  {author} {\bibinfo {author} {\bibfnamefont {D.~E.}\ \bibnamefont
  {Sheehy}}\ and\ \bibinfo {author} {\bibfnamefont {J.}~\bibnamefont
  {Schmalian}},\ }\href {\doibase 10.1103/PhysRevLett.99.226803} {\bibfield
  {journal} {\bibinfo  {journal} {Phys. Rev. Lett.}\ }\textbf {\bibinfo
  {volume} {99}},\ \bibinfo {eid} {226803} (\bibinfo {year}
  {2007})}\BibitemShut {NoStop}%
\bibitem [{\citenamefont {Nomura}\ \emph {et~al.}(2007)\citenamefont {Nomura},
  \citenamefont {Koshino},\ and\ \citenamefont {Ryu}}]{2007-nomura-fk}%
  \BibitemOpen
  \bibfield  {author} {\bibinfo {author} {\bibfnamefont {K.}~\bibnamefont
  {Nomura}}, \bibinfo {author} {\bibfnamefont {M.}~\bibnamefont {Koshino}}, \
  and\ \bibinfo {author} {\bibfnamefont {S.}~\bibnamefont {Ryu}},\ }\href
  {\doibase 10.1103/PhysRevLett.99.146806} {\bibfield  {journal} {\bibinfo
  {journal} {Phys. Rev. Lett.}\ }\textbf {\bibinfo {volume} {99}},\ \bibinfo
  {pages} {146806} (\bibinfo {year} {2007})}\BibitemShut {NoStop}%
\bibitem [{\citenamefont {McCann}\ \emph {et~al.}(2006)\citenamefont {McCann},
  \citenamefont {Kechedzhi}, \citenamefont {Fal'ko}, \citenamefont {Suzuura},
  \citenamefont {Ando},\ and\ \citenamefont {Altshuler}}]{2006-mccann-kx}%
  \BibitemOpen
  \bibfield  {author} {\bibinfo {author} {\bibfnamefont {E.}~\bibnamefont
  {McCann}}, \bibinfo {author} {\bibfnamefont {K.}~\bibnamefont {Kechedzhi}},
  \bibinfo {author} {\bibfnamefont {V.~I.}\ \bibnamefont {Fal'ko}}, \bibinfo
  {author} {\bibfnamefont {H.}~\bibnamefont {Suzuura}}, \bibinfo {author}
  {\bibfnamefont {T.}~\bibnamefont {Ando}}, \ and\ \bibinfo {author}
  {\bibfnamefont {B.~L.}\ \bibnamefont {Altshuler}},\ }\href {\doibase
  10.1103/PhysRevLett.97.146805} {\bibfield  {journal} {\bibinfo  {journal}
  {Phys. Rev. Lett.}\ }\textbf {\bibinfo {volume} {97}},\ \bibinfo {pages}
  {146805} (\bibinfo {year} {2006})}\BibitemShut {NoStop}%
\bibitem [{\citenamefont {Fradkin}(1986{\natexlab{a}})}]{1986-fradkin-yq}%
  \BibitemOpen
  \bibfield  {author} {\bibinfo {author} {\bibfnamefont {E.}~\bibnamefont
  {Fradkin}},\ }\href {\doibase 10.1103/PhysRevB.33.3257} {\bibfield  {journal}
  {\bibinfo  {journal} {Phys. Rev. B}\ }\textbf {\bibinfo {volume} {33}},\
  \bibinfo {pages} {3257} (\bibinfo {year} {1986}{\natexlab{a}})}\BibitemShut
  {NoStop}%
\bibitem [{\citenamefont {Fradkin}(1986{\natexlab{b}})}]{1986-fradkin-rt}%
  \BibitemOpen
  \bibfield  {author} {\bibinfo {author} {\bibfnamefont {E.}~\bibnamefont
  {Fradkin}},\ }\href {\doibase 10.1103/PhysRevB.33.3263} {\bibfield  {journal}
  {\bibinfo  {journal} {Phys. Rev. B}\ }\textbf {\bibinfo {volume} {33}},\
  \bibinfo {pages} {3263} (\bibinfo {year} {1986}{\natexlab{b}})}\BibitemShut
  {NoStop}%
\bibitem [{\citenamefont {Baym}\ and\ \citenamefont
  {Kadanoff}(1961)}]{1961-baym-yq}%
  \BibitemOpen
  \bibfield  {author} {\bibinfo {author} {\bibfnamefont {G.}~\bibnamefont
  {Baym}}\ and\ \bibinfo {author} {\bibfnamefont {L.~P.}\ \bibnamefont
  {Kadanoff}},\ }\href {\doibase 10.1103/PhysRev.124.287} {\bibfield  {journal}
  {\bibinfo  {journal} {Phys. Rev.}\ }\textbf {\bibinfo {volume} {124}},\
  \bibinfo {pages} {287} (\bibinfo {year} {1961})}\BibitemShut {NoStop}%
\bibitem [{\citenamefont {Lee}\ and\ \citenamefont
  {Ramakrishnan}(1985)}]{1985-lee-uq}%
  \BibitemOpen
  \bibfield  {author} {\bibinfo {author} {\bibfnamefont {P.~A.}\ \bibnamefont
  {Lee}}\ and\ \bibinfo {author} {\bibfnamefont {T.~V.}\ \bibnamefont
  {Ramakrishnan}},\ }\href {\doibase 10.1103/RevModPhys.57.287} {\bibfield
  {journal} {\bibinfo  {journal} {Rev. Mod. Phys.}\ }\textbf {\bibinfo {volume}
  {57}},\ \bibinfo {pages} {287} (\bibinfo {year} {1985})}\BibitemShut
  {NoStop}%
\bibitem [{\citenamefont {Altland}\ and\ \citenamefont
  {Simons.}(2010)}]{2010-altland-vn}%
  \BibitemOpen
  \bibfield  {author} {\bibinfo {author} {\bibfnamefont {A.}~\bibnamefont
  {Altland}}\ and\ \bibinfo {author} {\bibfnamefont {B.~D.}\ \bibnamefont
  {Simons.}},\ }\href {\doibase 10.1017/CBO9780511789984.001} {\emph {\bibinfo
  {title} {Condensed Matter Field Theory}}},\ \bibinfo {edition} {2nd}\ ed.\
  (\bibinfo  {publisher} {Cambridge University Press},\ \bibinfo {year}
  {2010})\BibitemShut {NoStop}%
\bibitem [{\citenamefont {Ando}\ and\ \citenamefont
  {Nakanishi}(1998)}]{1998-ando-lr}%
  \BibitemOpen
  \bibfield  {author} {\bibinfo {author} {\bibfnamefont {T.}~\bibnamefont
  {Ando}}\ and\ \bibinfo {author} {\bibfnamefont {T.}~\bibnamefont
  {Nakanishi}},\ }\href {\doibase 10.1143/JPSJ.67.1704} {\bibfield  {journal}
  {\bibinfo  {journal} {J. Phys. Soc. Jpn.}\ }\textbf {\bibinfo {volume}
  {67}},\ \bibinfo {pages} {1704} (\bibinfo {year} {1998})}\BibitemShut
  {NoStop}%
\bibitem [{\citenamefont {Ando}\ \emph {et~al.}(1998)\citenamefont {Ando},
  \citenamefont {Nakanishi},\ and\ \citenamefont {Saito}}]{1998-ando-ly}%
  \BibitemOpen
  \bibfield  {author} {\bibinfo {author} {\bibfnamefont {T.}~\bibnamefont
  {Ando}}, \bibinfo {author} {\bibfnamefont {T.}~\bibnamefont {Nakanishi}}, \
  and\ \bibinfo {author} {\bibfnamefont {R.}~\bibnamefont {Saito}},\ }\href
  {\doibase 10.1143/JPSJ.67.2857} {\bibfield  {journal} {\bibinfo  {journal}
  {J. Phys. Soc. Jpn.}\ }\textbf {\bibinfo {volume} {67}},\ \bibinfo {pages}
  {2857} (\bibinfo {year} {1998})}\BibitemShut {NoStop}%
\bibitem [{\citenamefont {Bruus}\ and\ \citenamefont
  {Flensberg}(2004)}]{2004-bruus-lq}%
  \BibitemOpen
  \bibfield  {author} {\bibinfo {author} {\bibfnamefont {H.}~\bibnamefont
  {Bruus}}\ and\ \bibinfo {author} {\bibfnamefont {K.}~\bibnamefont
  {Flensberg}},\ }\href
  {http://global.oup.com/academic/product/many-body-quantum-theory-in-condensed-matter-physics-9780198566335}
  {\emph {\bibinfo {title} {Many-body Quantum Theory In Condensed Matter
  Physics: An Introduction}}},\ \bibinfo {edition} {1st}\ ed.\ (\bibinfo
  {publisher} {Oxford University Press},\ \bibinfo {year} {2004})\BibitemShut
  {NoStop}%
\bibitem [{\citenamefont {Edwards}(1958)}]{1958-edwards-fk}%
  \BibitemOpen
  \bibfield  {author} {\bibinfo {author} {\bibfnamefont {S.~F.}\ \bibnamefont
  {Edwards}},\ }\href {\doibase 10.1080/14786435808243244} {\bibfield
  {journal} {\bibinfo  {journal} {Philosophical Magazine}\ }\textbf {\bibinfo
  {volume} {3}},\ \bibinfo {pages} {1020} (\bibinfo {year} {1958})}\BibitemShut
  {NoStop}%
\bibitem [{\citenamefont {Forster}(1990)}]{1990-forster-qq}%
  \BibitemOpen
  \bibfield  {author} {\bibinfo {author} {\bibfnamefont {D.}~\bibnamefont
  {Forster}},\ }\href@noop {} {\emph {\bibinfo {title} {Hydrodynamic
  Fluctuations, Broken Symmetry, And Correlation Functions}}}\ (\bibinfo
  {publisher} {Westview Press},\ \bibinfo {year} {1990})\BibitemShut {NoStop}%
\bibitem [{Note1()}]{Note1}%
  \BibitemOpen
  \bibinfo {note} {The generalized Laplace transform of a function $f(t)$ is
  defined via $$\protect \mathaccentV {tilde}07E{f}(z) = \DOTSI \intop
  \ilimits@ _{-\infty }^{\infty } \protect \frac {d\omega }{2\pi }\protect
  \frac {\protect \mathaccentV {hat}05E{f}(\omega )}{z - \omega },$$ where
  $\protect \mathaccentV {hat}05E{f}(\omega )$ is the usual Fourier transform
  of $f(t)$. With this definition, $\protect \mathaccentV {tilde}07E{f}(\omega
  \pm i 0)$ is the Fourier transform of $\mp i\Theta (\pm t)f(t)$, $\Theta $
  denoting the Heaviside step function.}\BibitemShut {Stop}%
\bibitem [{Note2()}]{Note2}%
  \BibitemOpen
  \bibinfo {note} {Scattering processes will effectively mix states whose
  energies are separated by the inverse state lifetime, i.e, the energy
  `linewidth'.}\BibitemShut {Stop}%
\bibitem [{Note3()}]{Note3}%
  \BibitemOpen
  \bibinfo {note} {This arises from the delta function in momentum space $
  \mathopen {\setbox \z@ \hbox {\frozen@everymath \@emptytoks \mathsurround \z@
  $\nulldelimiterspace \z@ \left <\vcenter to\@ne \big@size {}\right .$}\box
  \z@ } U_{\protect \bm {k}\protect \bm {q}}U_{\protect \bm {p}\protect \bm
  {l}}\mathclose {\setbox \z@ \hbox {\frozen@everymath \@emptytoks
  \mathsurround \z@ $\nulldelimiterspace \z@ \left >\vcenter to\@ne \big@size
  {}\right .$}\box \z@ } \propto \delta (\protect \bm {k}+\protect \bm {p}-
  \protect \bm {q}- \protect \bm {l}) $.}\BibitemShut {Stop}%
\bibitem [{\citenamefont {Vollhardt}\ and\ \citenamefont
  {W\"{o}lfle}(1980)}]{1980-vollhardt-fk}%
  \BibitemOpen
  \bibfield  {author} {\bibinfo {author} {\bibfnamefont {D.}~\bibnamefont
  {Vollhardt}}\ and\ \bibinfo {author} {\bibfnamefont {P.}~\bibnamefont
  {W\"{o}lfle}},\ }\href {\doibase 10.1103/PhysRevB.22.4666} {\bibfield
  {journal} {\bibinfo  {journal} {Phys. Rev. B}\ }\textbf {\bibinfo {volume}
  {22}},\ \bibinfo {pages} {4666} (\bibinfo {year} {1980})}\BibitemShut
  {NoStop}%
\bibitem [{\citenamefont {Abrikosov}\ \emph {et~al.}(1963)\citenamefont
  {Abrikosov}, \citenamefont {Gorkov},\ and\ \citenamefont
  {Dzyaloshinskii}}]{1963-abrikosov-ly}%
  \BibitemOpen
  \bibfield  {author} {\bibinfo {author} {\bibfnamefont {A.~A.}\ \bibnamefont
  {Abrikosov}}, \bibinfo {author} {\bibfnamefont {L.~P.}\ \bibnamefont
  {Gorkov}}, \ and\ \bibinfo {author} {\bibfnamefont {I.}~\bibnamefont
  {Dzyaloshinskii}},\ }\href {http://lccn.loc.gov/75017174} {\emph {\bibinfo
  {title} {Methods of quantum field theory in statistical physics}}}\ (\bibinfo
   {publisher} {Dover Publications},\ \bibinfo {address} {New York},\ \bibinfo
  {year} {1963})\BibitemShut {NoStop}%
\bibitem [{Note4()}]{Note4}%
  \BibitemOpen
  \bibinfo {note} {Using the formula \protect \url
  {http://dlmf.nist.gov/18.18##E9}.}\BibitemShut {Stop}%
\bibitem [{Note5()}]{Note5}%
  \BibitemOpen
  \bibinfo {note} {We are grateful to Tom Witten for pointing this
  out.}\BibitemShut {Stop}%
\bibitem [{Note6()}]{Note6}%
  \BibitemOpen
  \bibinfo {note} {The convergence of the saddle point estimate to the actual
  value of the integral is very slow as $t\to \infty $ and has not been
  included in Figure~\ref {fig-chargevstime}.}\BibitemShut {Stop}%
\bibitem [{\citenamefont {Martin}\ \emph {et~al.}(2007)\citenamefont {Martin},
  \citenamefont {Akerman}, \citenamefont {Ulbricht}, \citenamefont {Lohmann},
  \citenamefont {Smet}, \citenamefont {von Klitzing},\ and\ \citenamefont
  {Yacoby}}]{2007-martin-rt}%
  \BibitemOpen
  \bibfield  {author} {\bibinfo {author} {\bibfnamefont {J.}~\bibnamefont
  {Martin}}, \bibinfo {author} {\bibfnamefont {N.}~\bibnamefont {Akerman}},
  \bibinfo {author} {\bibfnamefont {G.}~\bibnamefont {Ulbricht}}, \bibinfo
  {author} {\bibfnamefont {T.}~\bibnamefont {Lohmann}}, \bibinfo {author}
  {\bibfnamefont {J.~H.}\ \bibnamefont {Smet}}, \bibinfo {author}
  {\bibfnamefont {K.}~\bibnamefont {von Klitzing}}, \ and\ \bibinfo {author}
  {\bibfnamefont {A.}~\bibnamefont {Yacoby}},\ }\href {\doibase
  10.1038/nphys781} {\bibfield  {journal} {\bibinfo  {journal} {Nat Phys}\
  }\textbf {\bibinfo {volume} {4}},\ \bibinfo {pages} {144} (\bibinfo {year}
  {2007})}\BibitemShut {NoStop}%
\end{thebibliography}
\end{document}